\documentclass[reprint,
amsmath,
amssymb,
 aps,
prb]{revtex4-1}
\usepackage{chemformula} 
\usepackage{bm}
\usepackage{amssymb}
\usepackage{float}
\usepackage{subcaption}
\usepackage[english]{babel}
\usepackage{tabularx,ragged2e,booktabs,caption}
\usepackage{amsmath}

\usepackage{hyperref}
\usepackage{algpseudocode}
\usepackage{mathtools}
\usepackage{mhchem}

\DeclarePairedDelimiter\norm{\lVert}{\rVert}%

\DeclareMathOperator*{\argmin}{argmin}

\newcommand{\grad}{\vec{\nabla}}
\newcommand{\laplacian}{\nabla^2}

\newcommand{\change}{}

\begin{document}

\author{Tingtao Zhou}
\affiliation{Massachusetts Institute of Technology, Department of Physics}
\author{Mohammad Mirzadeh}
\affiliation{Massachusetts Institute of Technology, Department of Chemical Engineering}
\author{Roland J.-M. Pellenq}
\affiliation{The MIT / CNRS / Aix-Marseille University Joint Laboratory,
"Multi-Scale Materials Science for Energy and Environment" and Massachusetts Institute of Technology, Department of Civil and Environmental Engineering}
\author{Martin Z. Bazant}
\affiliation{Massachusetts Institute of Technology, Department of Chemical Engineering}\affiliation{Massachusetts Institute of Technology, Department of  Mathematics}

\title{Freezing point depression and freeze-thaw damage by nano-fluidic salt trapping}





\begin{abstract}
A remarkable variety of organisms and wet materials are able to endure temperatures far below the freezing point of bulk water.  Cryo-tolerance in biology is usually attributed to ``anti-freeze" proteins, and yet massive supercooling ($< -40^\circ$C) is also possible in porous media containing only simple aqueous electrolytes. For concrete pavements, the common wisdom is that freeze-thaw (FT) damage results from the expansion of water upon freezing, but this cannot explain the large pressures ($> 10$~MPa) required to damage concrete, the observed correlation between pavement damage and de-icing salts, or the \change{FT} damage of cement paste loaded with benzene (which contracts upon freezing).  In this work, we propose a different mechanism -- nanofluidic salt trapping -- which can explain the observations, using simple mathematical models of dissolved ions confined 
between growing ice and charged pore surfaces.  
\change{When the transport timescale for ions through charged pore space is prolonged, ice formation in confined} pores causes enormous disjoining pressures via the ions \change{rejected from the ice core}, until their removal by precipitation or surface adsorption at a lower temperatures releases the pressure and allows complete freezing.  
The theory is able to predict the non-monotonic salt-concentration dependence of \change{FT} damage in concrete and provides some hint to better understand the origins of cryo-tolerance from a physical chemistry perspective.
\end{abstract}


\maketitle

\section{Introduction}



The durability of wet porous materials against freeze-thaw (FT) damage is critical in many areas of science and engineering.    In biology, it is a matter of life and death.    Living cells must somehow maintain a liquid state within the cellular membrane during winter~\citep{guy1990cold, saragusty2011current, storey2004physiology}, while avoiding anoxia due to external ice encasement \citep{andrews1996plants}.
Various anti-freeze proteins have been identified in cryo-tolerant animals, and  cryo-protectant chemicals have been used for cryo-preservation and {\it in vitro} fertilization~\citep{abdelhafez2010slow, saragusty2011current, wakchaure2015review, rienzi2017oocyte}. 
In addition, the complex thermodynamics of supercooled water could play a role. Even in bulk water, deep supercooling can lead to multiple metastable disordered states~\cite{mishima1998relationship,tulk2002structural,debenedetti2003supercooled}.  Phase transitions under nano-confinement~\cite{gelb1999phase,limmer2012phase} can lead to exotic new phases, as well as modified ice nucleation, in both experiments~\cite{morishige2003x,nosonovsky2008phase,agrawal2017observation} and molecular simulations ~\cite{bonnaud2010molecular,cerveny2016confined} of water in nanopores.   

In engineering, the most familiar example of FT damage is the fracture of concrete pavements during the winter \citep{farnam2014measuring},  commonly attributed to the expansion of water transforming to ice within the pores~\citep{cai1998freeze,PCAwebsitefreezethaw}.  However, this contradicts the observation that FT damage occurs when cement is loaded with benzene~\citep{beaudoin1974mechanism}, a normal liquid that shrinks upon freezing. 
Recent experiments have challenged the prevailing hypothesis that FT damage is directly caused by solid phase transformations, not only ice formation~\citep{farnam2014measuring,cai1998freeze}, but also salt crystallization~\citep{winkler1972crystallization, steiger2005crystal}. 
Interestingly, there is strong correlation between FT damage and the use of de-icing salts on concrete pavements~\citep{farnam2014acoustic}, which are often less durable than concrete structures without salt exposure in the same cold climates. Moreover, FT damage only occurs when the water saturation level exceeds a critical value~\citep{li2011water}. 
Previous models of ``frost heave'' (see e.g. ref~\cite{wettlaufer2006premelting}) achieved some successes in explaining the deformation of saturated soils due to the dynamics of pre-melted liquid and its coupling with the solid. However, the applicability of these theories to hardened cement is questionable, due to its much higher stiffness compared to capillary stresses~\cite{zhou2019capillary,zhou2019multiscale}. 
In summary, despite the societal importance of FT damage in cement, a physics-based theory has not yet been developed that can predict the enormous pressures required ($>10$ MPa), as well as all puzzling observations above.

\begin{figure*}
\centering
\includegraphics[width=1\linewidth]{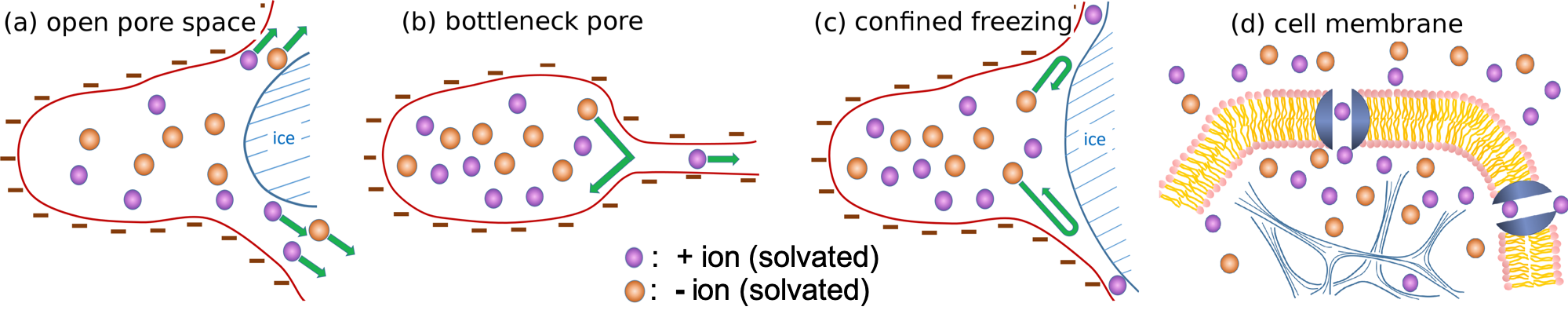}
\caption{
\label{fig:trap}
Physical picture of nanofluidic salt trapping. (a) In an open pore, where ions and water molecules can easily exchange with a nearby reservoir, no significant pressure or freezing point depression is predicted. (b) Nanoscale bottlenecks, especially in poorly connected porous networks, with charged surfaces can significantly hinder this exchange by size or charge exclusion of the co-ions, and counter-ions are forced to stay to maintain charge neutrality.  (c) Once ice nucleates, even an initially open pore will eventually trap a nanoscale thin film of supercooled, concentrated electrolyte near the charged surface,  until surface ion condensation or solid salt precipitation occurs. (d) In biological cells, the charged cytoskeleton (indicated by fibers) could enable such passive nanofluidic salt trapping, while further active control of water and ion flux across the cell membrane is performed by ion channels and pumps.  In all cases, nanofluidic salt trapping can lead to dramatic supercooling and, once ice nucleates, severe damage to the solid matrix.  }
\end{figure*}

In this article, we develop a predictive theory of freezing-point depression and FT damage in charged porous media, based on a simple new mechanism sketched in Figure ~\ref{fig:trap}: nanofluidic salt trapping. 
It is well known in colloid science that, when  two charged surfaces are separated by a liquid electrolyte, the crowding of ions in solution results \change{in} large repulsive forces whenever the electric double layers overlap, at the scale of the Debye screening length (1-100nm in water).  This ``disjoining pressure"  is responsible for the stabilization of colloidal dispersions in aqueous electrolytes~\cite{lyklema_book_vol2}, surface forces in clays and other porous media~\cite{israelachvili_book}, and the electrostatic properties of membranes~\cite{andelman1995_PBforces}. Disjoining pressure has been successfully modeled by the Poisson-Boltzmann mean-field theory for solutions of monovalent ions, and extensions are available to describe correlation effects involving multivalent ions~\cite{pellenq2004does,pellenq1997electrostatic}. Here we treat the disjoining pressure between ice core and the charge pore surface with a mean-field approximation. Although the physics of electrolyte freezing under confinement has been considered for nanoporous materials~\cite{alba2006effects}, we propose that nano-fluidic salt trapping is the key mechanism for large supercooling and  FT damage in cement and other charged nano-porous materials. This physical picture is consistent with all the available experimental evidence for concrete.


\subsection{Physical Picture}
Consider a heterogeneous porous material saturated with liquid and subjected to continuously decreasing temperatures.  As in most organisms and construction materials, suppose that the pore surfaces and suspended materials are hydrophilic and charged ~\cite{jan1973role, visser1973adhesion, giese1973interlayer, mclaughlin1977electrostatic, swartzen1974surface, allen2007composition, pellenq2009realistic}, e.g. by the dissociation of surface functional groups or the adsorption of charged species. The large capillary pores ($>$5~nm) are typically also filled with water, but can be replaced with other fluids such as benzene. However, the small ``gel'' pores ($\sim$1-5~nm) are always filled with liquid water due to the strong surface charge and hydrophilicity even in benzene-loaded cement samples.  Importantly, the liquid must contain dissolved salts, possibly at low concentration, as well as excess counter-ions to screen the pore surface charges and preserve overall electroneutrality.  Ions in solution mediate surface forces~\cite{israelachvili_book}, which play a crucial role in the mechanical properties of concrete~\cite{pellenq1997electrostatic, delville1998correlations, jellander1988attractive, stankovich1996electrical, valleau1980primitive, pellenq2004does} and the function of biological systems. In most cases, the ions are assumed to have negligible solubility in the frozen solid, as is the case with pure ice.

Freezing begins in the larger ``macropores" ($> 100$ nm), where bulk water easily transforms to ice, slightly below the thermodynamic melting point of the solution, which may be depressed from that of the pure solvent by the dissolved salt and any anti-freeze solutes.  This bulk ice can form by homogeneous nucleation, spinodal decomposition, or (most likely) by heterogeneous nucleation on impurities.  Regardless of its origin, the advancing ice rejects ions, causing the salt concentration to rise in the nearby, increasingly confined liquid electrolyte.

What happens next depends on the degree of supercooling, the surface charge and importantly the pore connectivity.  As shown in Figure ~\ref{fig:trap}(a), even after partial freezing, an individual pore may remain open, allowing ions and water molecules to exchange freely with a reservoir of bulk solution via a percolating liquid path to neighboring unfrozen pores or an external bath
~\citep{torquato2013random, van1975pore, quiblier1984new, pinson2018inferring,gu2018microscopic}.
In this scenario, the liquid electrolyte and any solid ice within the pore remain in quasi-equilibrium with the bulk reservoir at constant chemical potential.   The connected path  to the reservoir may pass through liquid-saturated pores, or partially frozen pores with sufficiently thick liquid films to allow unhindered transport.

As freezing proceeds, many ions and water molecules will inevitably be trapped out of {\it global} equilibrium, although still in {\it local} quasi-equilibrium within each nanoscale pore.  The simplest case is that of a pore connected to external reservoir only via a bottleneck, sketched in Fig.~\ref{fig:trap}(b). Water molecules that are not closely associated with ions can still go through the bottleneck, with a possibly different viscosity. The bottleneck may block solvated ions (with their solvation shells) from passing by steric hindrance or charge exclusion. Even if some solvated ions can diffuse through a given bottleneck, their electrokinetic transport rate may be too slow to allow many to escape prior to more complete freezing ~\citep{biesheuvel2016analysis, peters2016analysis, catalano2016theory}. Such slow ion transport may be enhanced by long, tortuous pathways through a series of bottlenecks \citep{yossifon2006electro, pennathur2005electrokinetic, chang2012nanoscale, yossifon2009nonlinear} and compounded by a large volume of micropores, effectively cut off from the macropores with insufficient time for salt release, in materials of low pore-space accessivity~\citep{gu2018microscopic}.
Even in relatively well connected porous structures, nanofluidic salt trapping can also result from bottlenecks created by the advancing ice, as shown in Figure ~\ref{fig:trap}(c), \change{where the larger open pore on the right side freezes almost completely first}. \change{Due to the surface hydrophilicity}, a supercooled liquid film often remains between the \change{pore surface and the ice core} prior to complete freezing~\citep{gelb1999phase,denoyel2002simple,bonnaud2010molecular}, \change{which is now the only pathway for water and ions in the smaller pore shown on the left side}. As temperature \change{decreases further, ice formation starts in the left pore, but transport of solvated ions through the thin liquid film is now slow, and the entropy of these confined ions builds up a pressure}. In biological cells, as shown in Figure ~\ref{fig:trap}(d), electrolytes are contained within the cell walls, and nanofluidic salt trapping is facilitated by the charged cytoskeleton and abundant charged macromolecules (including cryo-resistant proteins). Internal salt concentrations are also actively maintained by ion channels and pumps in the cell membrane \citep{hille2001ion}. 

\change{To quantitatively calculate the timescales of freezing and ion transport, one needs to solve a proper electrokinetic model of the 3D charged pore structure, with information of the tortuosity and connectivity in addition to the pore sizes. Here we focus on the asymptotic behavior of very long ion transport timescale v.s. freezing timescale, which hereafter referred to as the limit of trapped ions. This approximation of timescale separation is similar in essence to the adiabatic or Born-Oppenheimer approximation~\cite{born1927quantentheorie}, where the short time quasi-equilibrium is solved---as we show in the next sections---neglecting the slowly changing physics, in our case the ion transport.
The phenomenon of  ion trapping in charged nanochannels, while water remains free to diffuse and flow to a nearby reservoir or larger pore,  is well established in the field of nanofluidics and forms the basis for various devices, such as electro-osmotic micropumps\citep{zeng2001fabrication}, nano-fluidic diodes and bipolar transistors \citep{daiguji2005nanofluidic, yossifon2006electro, yossifon2009rectification}, and nanofluidic ion separators \citep{gillespie2013separation}.}

The supercooling of confined liquids can be greatly enhanced by the salt rejected by freezing, as the remaining solution becomes more concentrated inside a trapped freezing pore.  Large disjoining pressures are then produced in the very \change{concentrated} liquid solution and transmitted to the solid matrix, potentially causing damage.  

At sufficiently low temperatures, salt-enhanced supercooling and freeze-thaw \change{pressure} are relieved by the sudden precipitation of ions from the concentrated liquid, thus allowing complete freezing of the pores.  Ions may also be cleared by adsorption reactions on the pore surface, which regulate and neutralize the surface charge.  


\section{Theory}
\change{As mentioned above, under the assumption of separated timescales for ion transport and freezing, we approximate the dynamic problem as a quasi-equilibrium problem: in the limit of free ions, ion and water transport is much faster than freezing; in the other limit of trapped ions, ion transport is much slower than freezing. The solutions of both limits can be unified in the same quasi-equilibrium mean-field framework. Below we present details of these solutions.}

\change{The mean-field free energy for a liquid electrolyte and its frozen solid inside a charged pore can be described by}
\begin{equation}
\begin{split}
F_{tot} & = F_{liquid} + F_{solid} + F_{interface} \\
& = \int_{V_s} dV \left( \mu_s - \mu_l - \frac{\epsilon_{s}}{2} \norm{\grad \phi}^2 \right)  \\
& + \int_{V_l} dV \left[
g(\{c_i\}) + \rho\phi - \frac{\epsilon_l}{2} \norm{\grad \phi}^2  \right] \\
& +  \sum_{j=s, l, sl} \int_{S_{j}} dS \left( \gamma_j+ q_j\phi \right) \\
\end{split}
\label{eq:Fgen}
\end{equation}
where the integrations are performed over volumes of solid ($V_s$) and liquid ($V_l$) with permittivities $\epsilon_s$ and $\epsilon_l$, respectively, and over surfaces of the solid-liquid interface ($S_{sl}$), the liquid-pore interface ($S_l$) and the solid-pore interface ($S_s$), with corresponding surface charge densities, $q_{sl}$, $q_{l}$ and $q_{s}$ and interfacial tensions, $\gamma_{sl}$, $\gamma_l$ and $\gamma_s$; $\mu_s - \mu_l$ is the bulk chemical potential difference between solid and liquid phases; $-\grad \phi$ is the electric field; $g(\{c_i\})$ the non-electric part of homogeneous liquid electrolyte free energy;  $c_i$ the concentration of ion species $i$ having charge $z_i e$; and $\rho = \sum_i z_i e c_i$ the net charge density, assumed to be negligible in the solid phase.  We focus on situations of complete wetting by the liquid,  $\gamma_s-\gamma_l \gg \gamma_{sl}$, in which case we can neglect $S_s$ and assume $S_l$ covers the entire pore surface.

Setting $\delta F_{tot} / \delta \phi = 0$ for bulk and surface variations, we obtain Poisson's equation
\begin{equation}
\begin{split}
\epsilon_{l} \change{\laplacian} \phi = -\rho \ \ \mbox{ in } V_l \\
\epsilon_{s} \change{\laplacian} \phi = 0 \ \ \mbox{ in } V_s
\end{split}
\end{equation}
and electrostatic boundary conditions
\begin{equation}
\begin{split}
q_{sl} = (\epsilon_s \vec{E_s} - \epsilon_l \vec{E_l})\cdot \hat{n}_{ls} \ \mbox{ on } S_{sl} \\
q_{l} = \epsilon_{l} \grad\phi \cdot \hat{n}_{l} \ \mbox{ on } S_l
\end{split}
\end{equation}

The equilibrium state of liquid-solid coexistence is found by minimizing the total free energy with respect to the position and shape of the solid-liquid interface, $S_{sl}$. Here, we consider two cases: (1) an open pore where ions of species $i$ exchange freely with a reservoir of concentration $c_i^\infty$, or (2) a pore with trapped ions, whose total number is fixed by screening the pore surface charge in the liquid, prior to freezing, by the mechanisms shown in Fig. ~\ref{fig:trap}.   Importantly, we neglect the effects of volume changes due to the water/ice transformation, under the assumption that liquid water molecules (of size $\sim 3 \AA$) are mobile and small enough to escape the pore as freezing progresses, regardless of whether solvated ions are trapped.  In contrast to the common wisdom about freeze-thaw damage in pavements, this picture must also hold for well-connected hierarchical porous materials such as concrete.

The preceding thermodynamic framework for confined electrolyte phase transformations can be extended in various ways, e.g. to account for ion-ion correlations~\cite{bazant2011double} (especially involving multivalent ions), finite ion sizes~\cite{bazant2009towards} and hydration surface forces~\cite{bohinc2012poisson,brown2015emergence}, but here we focus on the simplest Poisson-Boltzmann mean-field theory~\cite{andelman1995_PBforces}, which suffices to predict the basic physics of freezing-point depression and material damage.
The homogeneous free energy is then given by the ideal gas entropy for point-like ions, $g_i=c_i[\ln(v_ic_i)-1]$, \change{with $v_i$ the molecular volume,} and the electrostatic potential in the liquid electrolyte is then given by  the Poisson-Boltzmann (PB) equation:
\begin{equation}
-\epsilon_l \change{\laplacian}\phi_{\change{l}} = \rho = \sum_i z_i e c_i \text{;}\quad
c_i = c_i^{\infty} e^{-\beta	z_i e \phi_l}
\label{eqn:PB}
\end{equation}
Since we focus on highly confined electrolyte liquid films, we set the relative \change{permittivity}, $\epsilon_l = 10 \epsilon_0$, to that of water near dielectric saturation at high charge density \citep{booth1951dielectric, aguilella2009dielectric}.

To assess the prevalence of nanofluidic salt trapping within Poisson-Boltzmann theory, the state of a bottleneck shown in Fig. ~\ref{fig:trap} can be estimated by comparing the double layer thickness $\lambda_D$ (or hydrated ion size $a$) inside with its radius $R$:
if $\lambda_D\sim R$ (or $a\gtrsim R$) then the double layer(s) span across and the bottleneck is approximated as ``closed" to ions, since freezing rate may exceed ion transport rate, given a high tortuousity of the pore network. If $\lambda_D\ll R$ (or $a\lesssim R$), then the channel may be viewed as open to ion exchange. For an initial salt concentration of $0.1~mol/L$ in a binary monovalent electrolyte, (with relative permittivity $\epsilon_r\sim 10$), we find $\lambda_D \sim \sqrt{\frac{4 \pi  \epsilon_l k_B T}{2c_0 e^2}} \sim 0.5~nm$.

\subsection{Symmetric pores}
In order to obtain analytical results, we consider isotropic electrolyte freezing in $d$ dimensions, where ice nucleates to form a plate ($d=1$), cylinder ($d=2$) or sphere ($d=3$) of radius $r$ within a pore of the same symmetry, whose surface is located at $x=R$.  The total pore volume is $V(d)r^d$, and $S(d) r^{d-1}$ the surface area of the ice core ($x<r$), surrounded by a liquid electrolyte shell ($r<x<R$). At thermodynamic equilibrium, the location $r^*$ of the solid-liquid interface is determined by minimizing the total free energy with respect to $r$, $\delta F_{tot} / \delta r = 0$:
\begin{equation}
r^* = \argmin_r F_{tot}(r)
\label{eqn:minimization-free-energy}
\end{equation}
which yields the equilibrium ice volume fraction, $\chi=(r^*/R)^d$.
Once $r^*$ is found, mechanical equilibrium
at the solid-liquid interface gives pressure of both phases, which is transmitted to the pore boundary
\begin{equation}
P = -\left(\frac{\partial F_{solid} }{\partial r}\right)_{r=r^*} = \left( \frac{\partial F_{liquid}}{\partial r} \right)_{r=r^*}
\label{eqn:mechanical-balance-pressure}
\end{equation}
The first equality describes the tendency to form more ice and hence expand its volume, while the second equality shows the free energy cost to squeeze the electrolyte, resisting the growth of ice.

For a symmetric pore, after freezing starts, the free energy of ice is given by
\begin{equation}
F_{ice} = \left( \mu_s - \mu_l \right) V(d) r^d,
\end{equation}
where $(\mu_s - \mu_l)$ is the Gibbs free energy change per volume for bulk water freezing, which can be calculated \citep{denoyel2002simple} using the Gibbs-Helmholtz relation, as shown in ref.~\cite{zhou2019freezingtheory}. In principle, the electric field energy of the ice core ($x<r$) depends on its shape and the electrostatic boundary conditions, but vanishes here by symmetry.
The interfacial energy is $F_{surface} = \gamma_{sl} S(d) r^{d-1}$, which gives rise to the Gibbs-Thomson \citep{gibbs1879equilibrium} effect of freezing point depression for confined pure water. The free energy of the electrolyte shell is given by
\begin{equation}
\begin{split}
& \frac{F_{elec}}{S(d)} = q_l \phi(R) R^{d-1} + \\
& \int_r^R x^{d-1} dx \left[
g(\{c_i\}) + \rho \phi - \frac{\epsilon_l}{2} \norm{\grad \phi}^2
\right]
\label{eqn:free-energy-electrolyte}
\end{split}
\end{equation}
where the first term is the electrostatic energy of surface charges, and the integrand takes the form given above for mean-field theory of point-like ions.  
To summarize, we are solving a free boundary problem where the liquid-ice boundary position $r$ is unknown beforehand.
We adopt a numerical algorithm to search for the $r$ that minimizes total free energy at a given temperature $T$, surface charge density $q$ and initial salt concentration $c_0$: 
\begin{enumerate}
    \item starting from $r=0$, compute the total free energy $F(0)$.
    \item increment $r$ by a small amount $dr$, compute the total free energy $F(r)$.\\
    when computing the total free energy at a given $r$ value, we always solve the Poisson-Boltzmann Eqn.~\ref{eqn:PB} to obtain the electric potential profile $\phi$, and insert into the integration of Eqn.~\ref{eqn:free-energy-electrolyte}.
    \item after sweep $r$ from 0 to the pore size $R$, find the minimum of $F$ and the corresponding $r$ gives the position of the quasi-equilibrium ice front.
\end{enumerate}


\begin{figure}
\centering
\includegraphics[width=0.4\textwidth]{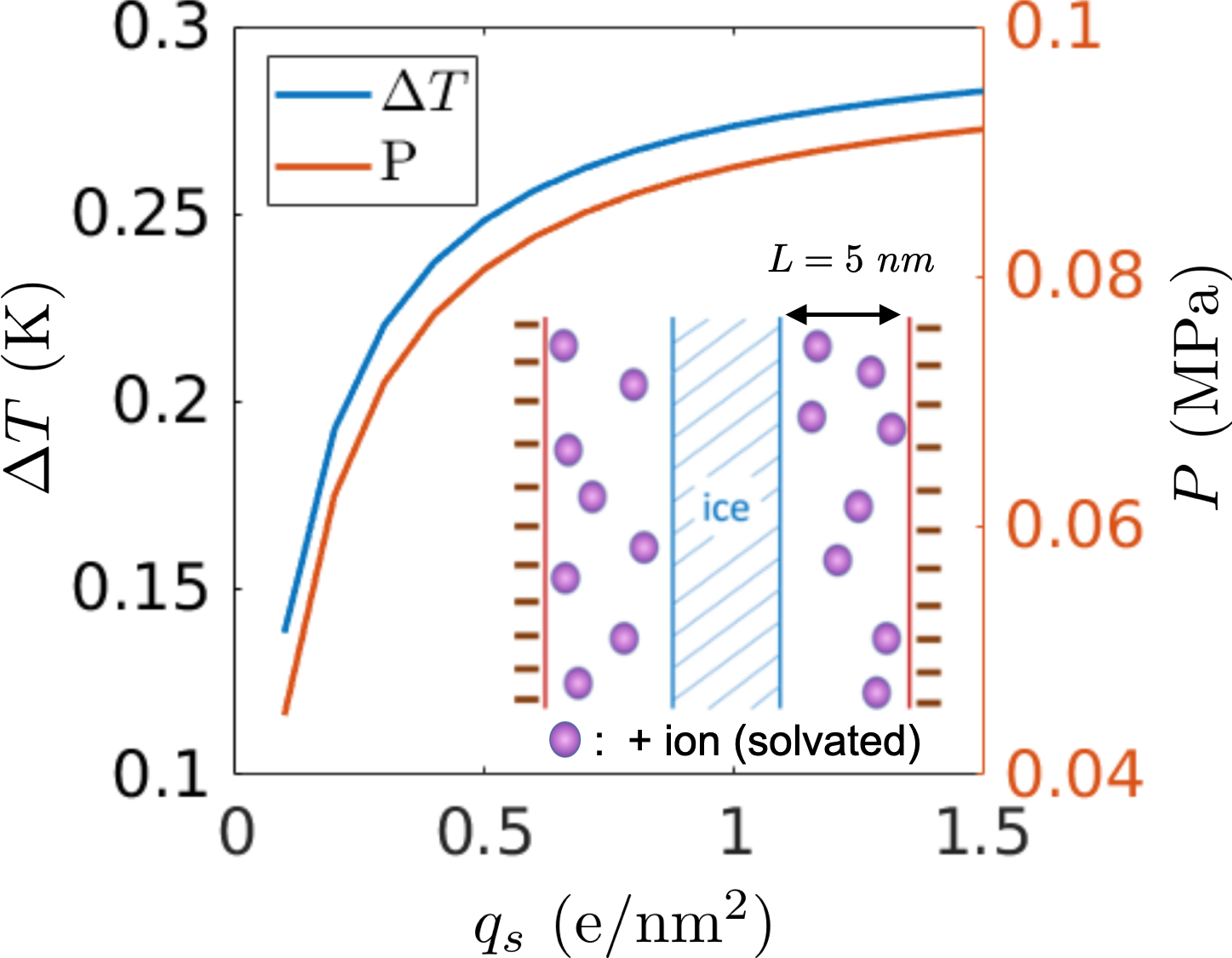}
\caption{
\label{fig:results-free-ion}
Electrolyte freezing and pressure generation in a parallel slit pore ($d=1$) with {\it free} ions exchanging with a reservoir. There is no effects of interfacial tension.
The freezing point depression, $\Delta T \sim 0.1$ K, and disjoining pressure, $P \sim 0.1$ MPa, are quite small, in the limit of one-component plasma of only counter-ions. In this case the total number of counter-ions is determined by the surface charge density only and does not depend on pore size. And the $\Delta T$ and $P$ only depends on the distance between ice front and the pore surface, which denoted by $L=5$~nm here.}
\end{figure}

\begin{figure}
\centering
\includegraphics[width=0.5\textwidth]{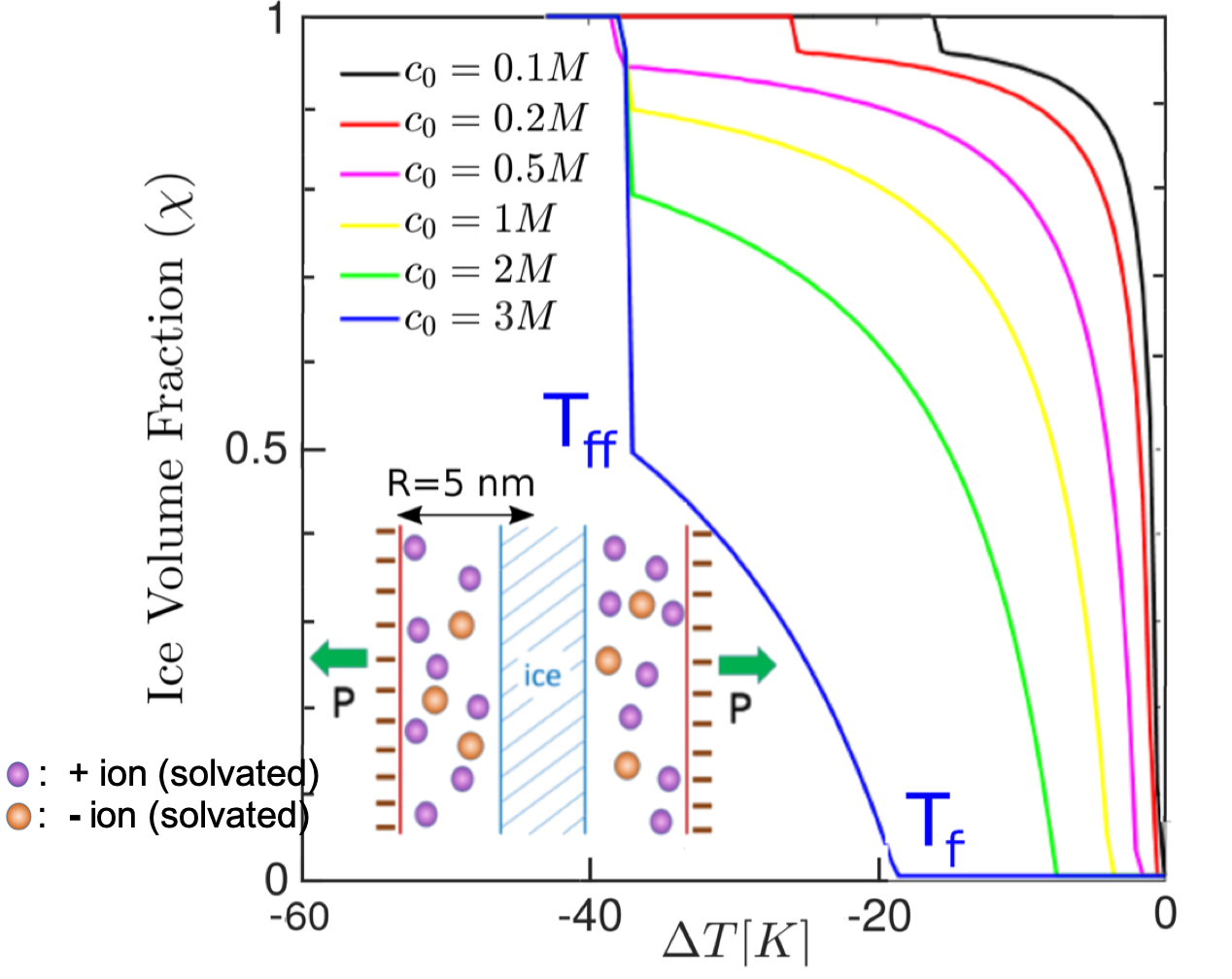}
\caption{
\label{fig:results-trapped-ion-T}
In contrast to Fig.\ref{fig:results-free-ion}, for a binary electrolyte with {\it trapped} ions, freezing-point depression as large as -40~K can occur. 
\change{The quasi-equilibrium approximation gives a continuous freezing temperature range marked by two $T$ values: the temperature to start freezing, $T_{f}$, and that of complete freezing of the pore $T_{ff}$,} when ions are removed by precipitation.
}
\end{figure}

\begin{figure}
\centering
\includegraphics[width=0.5\textwidth]{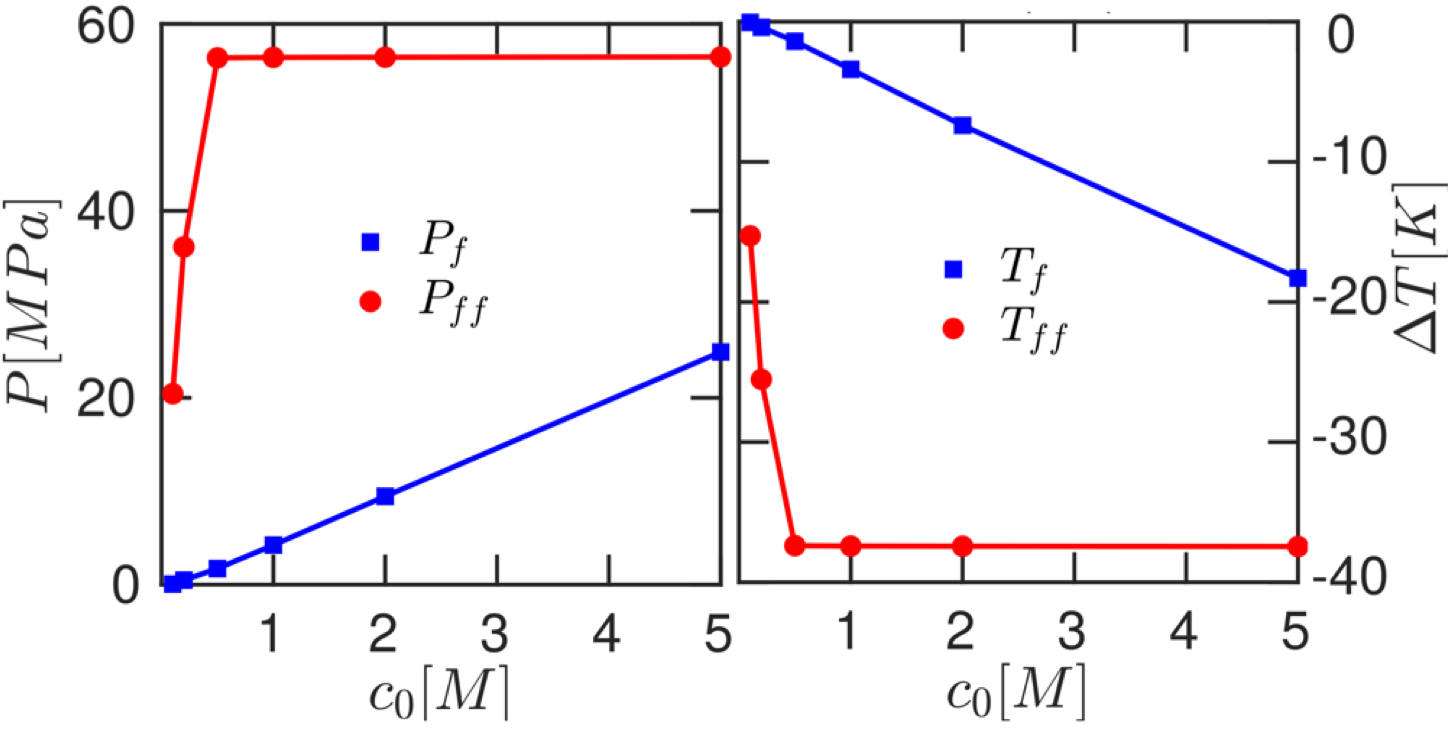}
\caption{
\label{fig:results-trapped-ion-P}
Large disjoining pressures up to $\sim$ 10~MPa occur during the freezing process, below the temperature to start freezing, $T_{f}$, and above that of complete freezing of the pore $T_{ff}$.
The range of pressure is marked by $P_f$ and $P_{ff}$, correspondingly. Blue and red dots are numerical results, while the solid lines connecting them are guiding the eyes.
}
\end{figure}

\subsection{Free ions}
As freezing proceeds in an open pore, where all ions can escape to a reservoir, the surface charge is eventually screened in a thin liquid film containing only counter-ions, which corresponds to one component plasma (OCP)~\citep{brush1966monte, baus1980statistical}.  
The Poisson-Boltzmann equation for the OCP can be integrated for symmetric pore shapes~\cite{zhou2019freezingtheory} to obtain the mean electrostatic potential. For a slit pore ($d=1$), we obtain
\change{ to the first order
\begin{equation}
\begin{split}
\vert\tilde{P}\vert & \approx \frac{4\pi \epsilon_l k_B T_0 Z^2}{q^2 e^2} \left\vert Q\frac{\Delta T}{T_0} 
\right\vert \\
\sqrt{\frac{\vert\tilde{P}\vert}{2}}  & \tan \left( \sqrt{\frac{\vert\tilde{P}\vert}{2}} \frac{R q e^2}{4\pi\epsilon_l k_B T} \right) = 2\pi
\end{split}
\end{equation}
}
where 
$Q$ is the latent heat of bulk water freezing, $T_0$ the bulk freezing point, and $\Delta T=T-T_0$ the freezing point depression. Notice that for OCP limit, the total amount of counter-ions does not depend on the pore size, but is simply determined by the surface charge density. Hence, the quasi-equilibrium solution only depends on the distance between the ice front and the pore surface, which we here denote as $L$.


Inserting typical values, we can estimate the freezing point depression in the slit pore as
$\Delta T \sim 0.1~K$
and the pressure as
$P_{elec} \sim 0.1 MPa$.
In this case, the freezing point is only depressed by $\lesssim 1$~K, and no significant pressure is generated, as shown in Fig.\ref{fig:results-free-ion}.  As shown in ref.~\cite{zhou2019freezingtheory}, the effects of ions in open cylindrical ($d=2$) or spherical ($d=3$) pores are even smaller than in a slit pore ($d=1$) and may often be neglected compared to the Gibbs-Thompson effect of interfacial tension in such curved geometries. In general, if excess salt ions (and water molecules) are free to escape the pore during freezing, then we expect very little freeze-thaw damage in a wet porous material.

\subsection{Trapped ions} 
The situation is completely different in the opposite limit, where all ions in the original liquid binary electrolyte remain trapped within the pore during freezing. Total ion number conservation is then imposed on the PB equations, $\int_r^R c_i S(d) x^{d-1}dx = N_i$, and significant freezing-point depression can be achieved.  The mathematical details can be found in a companion paper~\cite{zhou2019freezingtheory}, and here we focus on explaining the physical predictions of the theory. To separate the effect of curvature, here we focus on the slit symmetry ($d=1$).


First we consider a binary 1:1 liquid electrolyte freezing in a parallel slit pore ($d=1$). 
In this case, there is no effect of solid-liquid interfacial tension, as the interface area does not change as ice front advances (zero curvature).
As shown in Fig.\change{\ref{fig:results-trapped-ion-T}}, the freezing point is substantially decreased by increasing the initial salt concentration $c_0$ in the confined liquid. After freezing starts at temperature $T_{f}$, due to the resistance of the electrolyte, the equilibrium ice volume fraction $\chi$ monotonically increases as temperature decreases. The freezing process continues until the trapped ions are suddenly removed from the thin liquid film at the temperature of freezing finished $T_{ff}$, when the salt solubility limit is reached, and $\chi$ suddenly jumps to 1. The pore is completely frozen now. Complete freezing may also occur if the trapped ions are adsorbed on the pore surface, thereby neutralizing the surface charge (as shown below).

As shown in Figure \ref{fig:results-trapped-ion-P}, significant disjoining pressures ($\sim 10$ MPa for $R=5$ nm) can be generated by confined ions during the freezing process.   The pressures at the freezing start temperature $T_f$ and the complete freezing temperature $T_{ff}$ are labeled as $P_{f}$ and $P_{ff}$, respectively.  The disjoining pressure varies approximately linearly with temperature between these values during the freezing process in a slit pore.

\begin{figure}
\centering
\includegraphics[width=1\linewidth]{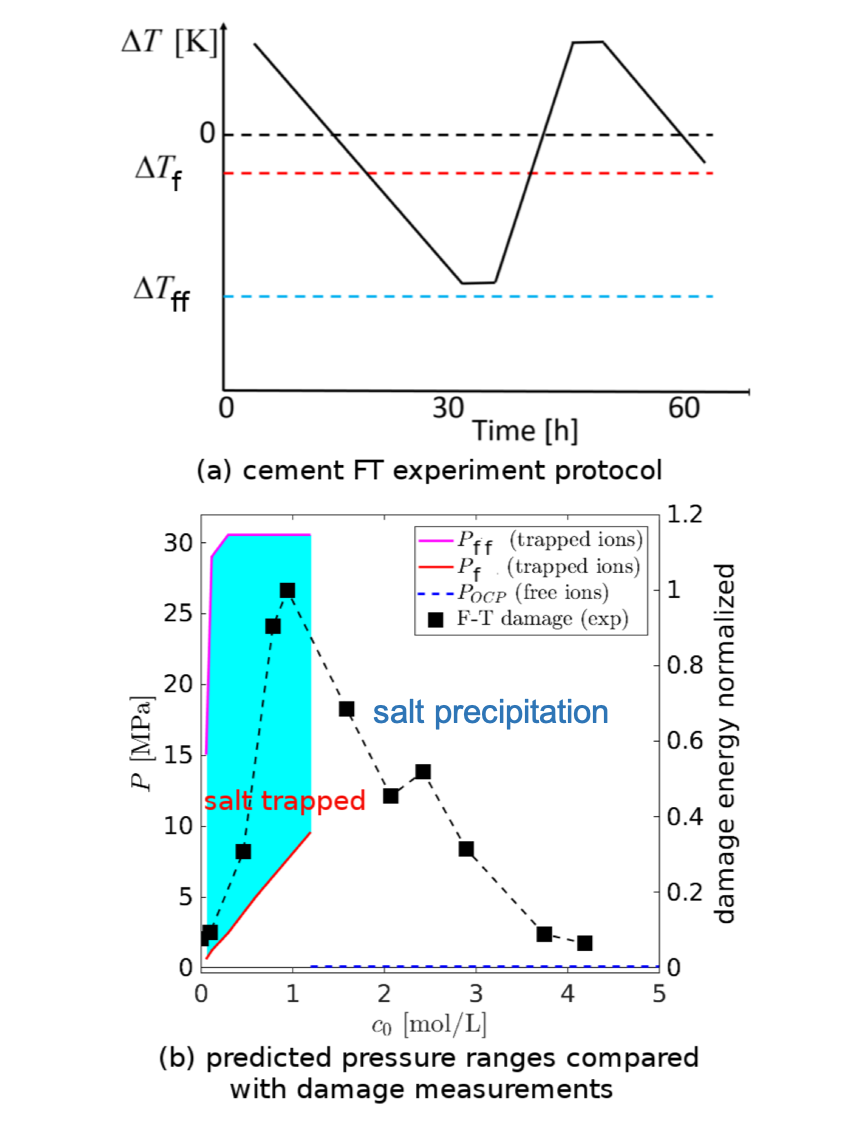}
\caption{
\label{fig:damage}
(a) typcial experiment protocol reproduced from \cite{farnam2014acoustic}. $T_f$ and $T_{ff}$ correspond to the temperature when ice formation initiates and solubility limit reached, as indicated in Fig.\ref{fig:results-trapped-ion-T}. (b) shaded area shows the pressure range after freezing starts in a trapped pore. $P_{f}$ and $P_{ff}$ correspond to the pressure when ice formation initiates and solubility limit reached, as indicated in Fig.\ref{fig:results-trapped-ion-P}. Dash line shows open pores with free ions, which is close to the horizontal line of 0. Data points show measured damage 
in cement paste FT experiment~\cite{farnam2014acoustic}, a non-monotonic function of salt concentration. Tensile strength of hardened cement paste is $\sim$ 3~MPa.}
\end{figure}

\subsection{Salt solubility limit and surface charge regulation} 
As ice volume fraction increases, salt concentration goes up. At some point the concentrated electrolyte will become saturated and salt will crystallize. The volume of salt crystal precipitate is neglected. The solubility equilibrium for 1:1 electrolyte ($\ce{M^+ + B^-}$) at saturation is
\change{
$K_{eq} = \frac{\ce{[M^+]}\ce{[B^-]}}{\ce{[MB]}} = \left(\frac{c_0^{sat}}{c_{solid}}\right)^2$. Here $c_{solid}$ is the concentration in solid crystal phase, which is typically regarded as constant 1.
Once $c_0^{sat}$, the saturated concentration of salt ions,} is reached the equilibrium position of ice front becomes thermodynamically unstable and all the liquid turned into solid phases of ice and salt crystal. In Fig.\ref{fig:results-trapped-ion-T}, all the curves at some point reach the solubility limit and undergo sudden crystallization, when ice volume fraction discontinuously jumps from $\chi<1$ to $\chi=1$. The pressure at this point is denoted as $P_{ff}$ in both Fig.\ref{fig:results-trapped-ion-P} and Fig.\ref{fig:damage}.
As opposed to the concept of ``crystallization pressure'' \citep{winkler1972crystallization, steiger2005crystal} that has been proposed to account for pressure and damage (under room temperature) in construction materials, here the pressure of the freezing pore is determined by thermodynamc equilibrium between freezing and precipitation, thus salt crystallization is merely a consequence, instead of the cause for pressure.

When the concentration of trapped ions is high enough, counter-ion  recombination with the surface charge becomes important. This effect can be included by a modified boundary condition for the PB equations, where the surface charge is computed self-consistently based on a charge regulation model~\citep{markovich2016charge} (see more details in~\cite{zhou2019freezingtheory}).

\section{Application to Concrete}
The predictions of this theory are semi-quantitatively consistent with experimental observations of freeze-thaw damage in cement. Below critical degree of water saturation, plenty of large pores remain open transport pathways for ions during freezing, hence no significant damage observed~\citep{li2011water}. The volume expansion of water during freezing is irrelevant in this theory, so it can also explain qualitatively similar results observed in freeze-thaw experimetns on cement samples loaded with benzene, which shrinks upon freezing~\cite{beaudoin1974mechanism}. The non-monotonic dependence of damage on $\ce{NaCl}$ concentration \citep{farnam2014acoustic} can be explained by crossover from salt trapping to channel opening though charge regulation, as shown in Fig.\ref{fig:damage}. A fully quantitative comparison requires the plasticity and fracture mechanics of the solid matrix due to these local high pressures, and the connectivity of the pores, which is currently a missing link. Also to quantify the transport timescales for ions as freezing proceeds, pore connectivity is key information. Nevertheless, to our knowledge for the first time this mechanism shows potential to encompass all these observations.

\section{Conclusion and discussions}
In this article, we present a theory of the freezing of electrolytes in charged porous media.  The key insight is that, if ions become trapped by the advancing ice front,  large disjoining pressures can cause material damage, until further supercooling triggers salt precipitation and complete freezing.   Freezing point depression, ice volume fraction and pressure are calculated using a simple mean-field theory.

Many extensions of the theory could be considered in future work. 
Ion correlations, including the strong-coupling limit~\citep{moreira2000strong, netz2001electrostatistics, vsamaj2011wigner}, can be introduced via higher order terms in Eqn.\ref{eqn:free-energy-electrolyte}, resulting in modified PB equations~\citep{bazant2011double}.
At larger length scales, models of interfacial instabilities leading to dendritic growth~\citep{hastings1996laplacian, mullins1963morphological, mullins1988stability} could be extended to account for electrokinetic phenomena in charged pores~\cite{bazant2003dynamics,mirzadeh2017electrokinetic}.
Here we always assume bulk phase of ice (the Ih phase) is formed, since the freezing conditions discussed here ($T>200~K$, $P<100~MPa$, $d<100$~nm) are not very extreme. Exotic phases of ice (non-Ih phases) are known to dominate under more extreme conditions \citep{dowell1960low, steytler1983neutron, mayer1987cubic, murray2005formation, moore2011cubic,  finney2002structures, tulk2002structural}. 
Salt ions can also affect the surface tension of ice-electrolyte interface, as well as other aspects of nucleation under confinement, described in a companion paper~\cite{zhou2019freezingtheory}.

As a first application to material durability, our theory is consistent with complex trends of freeze-thaw damage in hardened cement paste.
These predictions could influence industrial practices in road de-icing and pavement design.  The theory may also provide some perspective on the physics of cryo-tolerance and cryo-preservation in biological materials, which abound in electrolyte-soaked macromolecules, nanopores and membranes.

\section{acknowledgement}
This work is carried out with the support of the Concrete Sustainability Hub at MIT. The authors thank S. Yip, C. Qiao, J. Weiss, M. Pinson for useful discussions.




\bibliography{freezing-langmuir}

\begin{thebibliography}{89}%
\makeatletter
\providecommand \@ifxundefined [1]{%
 \@ifx{#1\undefined}
}%
\providecommand \@ifnum [1]{%
 \ifnum #1\expandafter \@firstoftwo
 \else \expandafter \@secondoftwo
 \fi
}%
\providecommand \@ifx [1]{%
 \ifx #1\expandafter \@firstoftwo
 \else \expandafter \@secondoftwo
 \fi
}%
\providecommand \natexlab [1]{#1}%
\providecommand \enquote  [1]{``#1''}%
\providecommand \bibnamefont  [1]{#1}%
\providecommand \bibfnamefont [1]{#1}%
\providecommand \citenamefont [1]{#1}%
\providecommand \href@noop [0]{\@secondoftwo}%
\providecommand \href [0]{\begingroup \@sanitize@url \@href}%
\providecommand \@href[1]{\@@startlink{#1}\@@href}%
\providecommand \@@href[1]{\endgroup#1\@@endlink}%
\providecommand \@sanitize@url [0]{\catcode `\\12\catcode `\$12\catcode
  `\&12\catcode `\#12\catcode `\^12\catcode `\_12\catcode `\%12\relax}%
\providecommand \@@startlink[1]{}%
\providecommand \@@endlink[0]{}%
\providecommand \url  [0]{\begingroup\@sanitize@url \@url }%
\providecommand \@url [1]{\endgroup\@href {#1}{\urlprefix }}%
\providecommand \urlprefix  [0]{URL }%
\providecommand \Eprint [0]{\href }%
\providecommand \doibase [0]{http://dx.doi.org/}%
\providecommand \selectlanguage [0]{\@gobble}%
\providecommand \bibinfo  [0]{\@secondoftwo}%
\providecommand \bibfield  [0]{\@secondoftwo}%
\providecommand \translation [1]{[#1]}%
\providecommand \BibitemOpen [0]{}%
\providecommand \bibitemStop [0]{}%
\providecommand \bibitemNoStop [0]{.\EOS\space}%
\providecommand \EOS [0]{\spacefactor3000\relax}%
\providecommand \BibitemShut  [1]{\csname bibitem#1\endcsname}%
\let\auto@bib@innerbib\@empty
\bibitem [{\citenamefont {Guy}(1990)}]{guy1990cold}%
  \BibitemOpen
  \bibfield  {author} {\bibinfo {author} {\bibfnamefont {C.~L.}\ \bibnamefont
  {Guy}},\ }\href@noop {} {\bibfield  {journal} {\bibinfo  {journal} {Annual
  review of plant biology}\ }\textbf {\bibinfo {volume} {41}},\ \bibinfo
  {pages} {187} (\bibinfo {year} {1990})}\BibitemShut {NoStop}%
\bibitem [{\citenamefont {Saragusty}\ and\ \citenamefont
  {Arav}(2011)}]{saragusty2011current}%
  \BibitemOpen
  \bibfield  {author} {\bibinfo {author} {\bibfnamefont {J.}~\bibnamefont
  {Saragusty}}\ and\ \bibinfo {author} {\bibfnamefont {A.}~\bibnamefont
  {Arav}},\ }\href@noop {} {\bibfield  {journal} {\bibinfo  {journal}
  {Reproduction}\ }\textbf {\bibinfo {volume} {141}},\ \bibinfo {pages} {1}
  (\bibinfo {year} {2011})}\BibitemShut {NoStop}%
\bibitem [{\citenamefont {Storey}\ and\ \citenamefont
  {Storey}(2004)}]{storey2004physiology}%
  \BibitemOpen
  \bibfield  {author} {\bibinfo {author} {\bibfnamefont {J.~M.}\ \bibnamefont
  {Storey}}\ and\ \bibinfo {author} {\bibfnamefont {K.~B.}\ \bibnamefont
  {Storey}},\ }in\ \href@noop {} {\emph {\bibinfo {booktitle} {Life in the
  frozen state}}}\ (\bibinfo  {publisher} {CRC press},\ \bibinfo {year}
  {2004})\ pp.\ \bibinfo {pages} {269--300}\BibitemShut {NoStop}%
\bibitem [{\citenamefont {Andrews}(1996)}]{andrews1996plants}%
  \BibitemOpen
  \bibfield  {author} {\bibinfo {author} {\bibfnamefont {C.}~\bibnamefont
  {Andrews}},\ }\href@noop {} {\bibfield  {journal} {\bibinfo  {journal}
  {Annals of Botany}\ }\textbf {\bibinfo {volume} {78}},\ \bibinfo {pages}
  {529} (\bibinfo {year} {1996})}\BibitemShut {NoStop}%
\bibitem [{\citenamefont {AbdelHafez}\ \emph {et~al.}(2010)\citenamefont
  {AbdelHafez}, \citenamefont {Desai}, \citenamefont {Abou-Setta},
  \citenamefont {Falcone},\ and\ \citenamefont
  {Goldfarb}}]{abdelhafez2010slow}%
  \BibitemOpen
  \bibfield  {author} {\bibinfo {author} {\bibfnamefont {F.~F.}\ \bibnamefont
  {AbdelHafez}}, \bibinfo {author} {\bibfnamefont {N.}~\bibnamefont {Desai}},
  \bibinfo {author} {\bibfnamefont {A.~M.}\ \bibnamefont {Abou-Setta}},
  \bibinfo {author} {\bibfnamefont {T.}~\bibnamefont {Falcone}}, \ and\
  \bibinfo {author} {\bibfnamefont {J.}~\bibnamefont {Goldfarb}},\ }\href@noop
  {} {\bibfield  {journal} {\bibinfo  {journal} {Reproductive biomedicine
  online}\ }\textbf {\bibinfo {volume} {20}},\ \bibinfo {pages} {209} (\bibinfo
  {year} {2010})}\BibitemShut {NoStop}%
\bibitem [{\citenamefont {Wakchaure}\ \emph {et~al.}(2015)\citenamefont
  {Wakchaure}, \citenamefont {Ganguly}, \citenamefont {Sharma}, \citenamefont
  {Praveen}, \citenamefont {Sharma},\ and\ \citenamefont
  {Mahajan}}]{wakchaure2015review}%
  \BibitemOpen
  \bibfield  {author} {\bibinfo {author} {\bibfnamefont {R.}~\bibnamefont
  {Wakchaure}}, \bibinfo {author} {\bibfnamefont {S.}~\bibnamefont {Ganguly}},
  \bibinfo {author} {\bibfnamefont {S.}~\bibnamefont {Sharma}}, \bibinfo
  {author} {\bibfnamefont {P.~K.}\ \bibnamefont {Praveen}}, \bibinfo {author}
  {\bibfnamefont {M.}~\bibnamefont {Sharma}}, \ and\ \bibinfo {author}
  {\bibfnamefont {T.}~\bibnamefont {Mahajan}},\ }\href@noop {} {\bibfield
  {journal} {\bibinfo  {journal} {Int. J. Phar. \& Biomedi. Rese}\ }\textbf
  {\bibinfo {volume} {2}},\ \bibinfo {pages} {11} (\bibinfo {year}
  {2015})}\BibitemShut {NoStop}%
\bibitem [{\citenamefont {Rienzi}\ \emph {et~al.}(2017)\citenamefont {Rienzi},
  \citenamefont {Gracia}, \citenamefont {Maggiulli}, \citenamefont {LaBarbera},
  \citenamefont {Kaser}, \citenamefont {Ubaldi}, \citenamefont {Vanderpoel},\
  and\ \citenamefont {Racowsky}}]{rienzi2017oocyte}%
  \BibitemOpen
  \bibfield  {author} {\bibinfo {author} {\bibfnamefont {L.}~\bibnamefont
  {Rienzi}}, \bibinfo {author} {\bibfnamefont {C.}~\bibnamefont {Gracia}},
  \bibinfo {author} {\bibfnamefont {R.}~\bibnamefont {Maggiulli}}, \bibinfo
  {author} {\bibfnamefont {A.~R.}\ \bibnamefont {LaBarbera}}, \bibinfo {author}
  {\bibfnamefont {D.~J.}\ \bibnamefont {Kaser}}, \bibinfo {author}
  {\bibfnamefont {F.~M.}\ \bibnamefont {Ubaldi}}, \bibinfo {author}
  {\bibfnamefont {S.}~\bibnamefont {Vanderpoel}}, \ and\ \bibinfo {author}
  {\bibfnamefont {C.}~\bibnamefont {Racowsky}},\ }\href@noop {} {\bibfield
  {journal} {\bibinfo  {journal} {Human reproduction update}\ }\textbf
  {\bibinfo {volume} {23}},\ \bibinfo {pages} {139} (\bibinfo {year}
  {2017})}\BibitemShut {NoStop}%
\bibitem [{\citenamefont {Mishima}\ and\ \citenamefont
  {Stanley}(1998)}]{mishima1998relationship}%
  \BibitemOpen
  \bibfield  {author} {\bibinfo {author} {\bibfnamefont {O.}~\bibnamefont
  {Mishima}}\ and\ \bibinfo {author} {\bibfnamefont {H.~E.}\ \bibnamefont
  {Stanley}},\ }\href@noop {} {\bibfield  {journal} {\bibinfo  {journal}
  {Nature}\ }\textbf {\bibinfo {volume} {396}},\ \bibinfo {pages} {329}
  (\bibinfo {year} {1998})}\BibitemShut {NoStop}%
\bibitem [{\citenamefont {Tulk}\ \emph {et~al.}(2002)\citenamefont {Tulk},
  \citenamefont {Benmore}, \citenamefont {Urquidi}, \citenamefont {Klug},
  \citenamefont {Neuefeind}, \citenamefont {Tomberli},\ and\ \citenamefont
  {Egelstaff}}]{tulk2002structural}%
  \BibitemOpen
  \bibfield  {author} {\bibinfo {author} {\bibfnamefont {C.}~\bibnamefont
  {Tulk}}, \bibinfo {author} {\bibfnamefont {C.}~\bibnamefont {Benmore}},
  \bibinfo {author} {\bibfnamefont {J.}~\bibnamefont {Urquidi}}, \bibinfo
  {author} {\bibfnamefont {D.}~\bibnamefont {Klug}}, \bibinfo {author}
  {\bibfnamefont {J.}~\bibnamefont {Neuefeind}}, \bibinfo {author}
  {\bibfnamefont {B.}~\bibnamefont {Tomberli}}, \ and\ \bibinfo {author}
  {\bibfnamefont {P.}~\bibnamefont {Egelstaff}},\ }\href@noop {} {\bibfield
  {journal} {\bibinfo  {journal} {Science}\ }\textbf {\bibinfo {volume}
  {297}},\ \bibinfo {pages} {1320} (\bibinfo {year} {2002})}\BibitemShut
  {NoStop}%
\bibitem [{\citenamefont {Debenedetti}(2003)}]{debenedetti2003supercooled}%
  \BibitemOpen
  \bibfield  {author} {\bibinfo {author} {\bibfnamefont {P.~G.}\ \bibnamefont
  {Debenedetti}},\ }\href@noop {} {\bibfield  {journal} {\bibinfo  {journal}
  {Journal of Physics: Condensed Matter}\ }\textbf {\bibinfo {volume} {15}},\
  \bibinfo {pages} {R1669} (\bibinfo {year} {2003})}\BibitemShut {NoStop}%
\bibitem [{\citenamefont {Gelb}\ \emph {et~al.}(1999)\citenamefont {Gelb},
  \citenamefont {Gubbins}, \citenamefont {Radhakrishnan},\ and\ \citenamefont
  {Sliwinska-Bartkowiak}}]{gelb1999phase}%
  \BibitemOpen
  \bibfield  {author} {\bibinfo {author} {\bibfnamefont {L.~D.}\ \bibnamefont
  {Gelb}}, \bibinfo {author} {\bibfnamefont {K.}~\bibnamefont {Gubbins}},
  \bibinfo {author} {\bibfnamefont {R.}~\bibnamefont {Radhakrishnan}}, \ and\
  \bibinfo {author} {\bibfnamefont {M.}~\bibnamefont {Sliwinska-Bartkowiak}},\
  }\href@noop {} {\bibfield  {journal} {\bibinfo  {journal} {Reports on
  Progress in Physics}\ }\textbf {\bibinfo {volume} {62}},\ \bibinfo {pages}
  {1573} (\bibinfo {year} {1999})}\BibitemShut {NoStop}%
\bibitem [{\citenamefont {Limmer}\ and\ \citenamefont
  {Chandler}(2012)}]{limmer2012phase}%
  \BibitemOpen
  \bibfield  {author} {\bibinfo {author} {\bibfnamefont {D.~T.}\ \bibnamefont
  {Limmer}}\ and\ \bibinfo {author} {\bibfnamefont {D.}~\bibnamefont
  {Chandler}},\ }\href@noop {} {\bibfield  {journal} {\bibinfo  {journal} {The
  Journal of chemical physics}\ }\textbf {\bibinfo {volume} {137}},\ \bibinfo
  {pages} {044509} (\bibinfo {year} {2012})}\BibitemShut {NoStop}%
\bibitem [{\citenamefont {Morishige}\ and\ \citenamefont
  {Iwasaki}(2003)}]{morishige2003x}%
  \BibitemOpen
  \bibfield  {author} {\bibinfo {author} {\bibfnamefont {K.}~\bibnamefont
  {Morishige}}\ and\ \bibinfo {author} {\bibfnamefont {H.}~\bibnamefont
  {Iwasaki}},\ }\href@noop {} {\bibfield  {journal} {\bibinfo  {journal}
  {Langmuir}\ }\textbf {\bibinfo {volume} {19}},\ \bibinfo {pages} {2808}
  (\bibinfo {year} {2003})}\BibitemShut {NoStop}%
\bibitem [{\citenamefont {Nosonovsky}\ and\ \citenamefont
  {Bhushan}(2008)}]{nosonovsky2008phase}%
  \BibitemOpen
  \bibfield  {author} {\bibinfo {author} {\bibfnamefont {M.}~\bibnamefont
  {Nosonovsky}}\ and\ \bibinfo {author} {\bibfnamefont {B.}~\bibnamefont
  {Bhushan}},\ }\href@noop {} {\bibfield  {journal} {\bibinfo  {journal}
  {Physical Chemistry Chemical Physics}\ }\textbf {\bibinfo {volume} {10}},\
  \bibinfo {pages} {2137} (\bibinfo {year} {2008})}\BibitemShut {NoStop}%
\bibitem [{\citenamefont {Agrawal}\ \emph {et~al.}(2017)\citenamefont
  {Agrawal}, \citenamefont {Shimizu}, \citenamefont {Drahushuk}, \citenamefont
  {Kilcoyne},\ and\ \citenamefont {Strano}}]{agrawal2017observation}%
  \BibitemOpen
  \bibfield  {author} {\bibinfo {author} {\bibfnamefont {K.~V.}\ \bibnamefont
  {Agrawal}}, \bibinfo {author} {\bibfnamefont {S.}~\bibnamefont {Shimizu}},
  \bibinfo {author} {\bibfnamefont {L.~W.}\ \bibnamefont {Drahushuk}}, \bibinfo
  {author} {\bibfnamefont {D.}~\bibnamefont {Kilcoyne}}, \ and\ \bibinfo
  {author} {\bibfnamefont {M.~S.}\ \bibnamefont {Strano}},\ }\href@noop {}
  {\bibfield  {journal} {\bibinfo  {journal} {Nature nanotechnology}\ }\textbf
  {\bibinfo {volume} {12}},\ \bibinfo {pages} {267} (\bibinfo {year}
  {2017})}\BibitemShut {NoStop}%
\bibitem [{\citenamefont {Bonnaud}\ \emph {et~al.}(2010)\citenamefont
  {Bonnaud}, \citenamefont {Coasne},\ and\ \citenamefont
  {Pellenq}}]{bonnaud2010molecular}%
  \BibitemOpen
  \bibfield  {author} {\bibinfo {author} {\bibfnamefont {P.~A.}\ \bibnamefont
  {Bonnaud}}, \bibinfo {author} {\bibfnamefont {B.}~\bibnamefont {Coasne}}, \
  and\ \bibinfo {author} {\bibfnamefont {R.~J.}\ \bibnamefont {Pellenq}},\
  }\href@noop {} {\bibfield  {journal} {\bibinfo  {journal} {Journal of
  Physics: Condensed Matter}\ }\textbf {\bibinfo {volume} {22}},\ \bibinfo
  {pages} {284110} (\bibinfo {year} {2010})}\BibitemShut {NoStop}%
\bibitem [{\citenamefont {Cerveny}\ \emph {et~al.}(2016)\citenamefont
  {Cerveny}, \citenamefont {Mallamace}, \citenamefont {Swenson}, \citenamefont
  {Vogel},\ and\ \citenamefont {Xu}}]{cerveny2016confined}%
  \BibitemOpen
  \bibfield  {author} {\bibinfo {author} {\bibfnamefont {S.}~\bibnamefont
  {Cerveny}}, \bibinfo {author} {\bibfnamefont {F.}~\bibnamefont {Mallamace}},
  \bibinfo {author} {\bibfnamefont {J.}~\bibnamefont {Swenson}}, \bibinfo
  {author} {\bibfnamefont {M.}~\bibnamefont {Vogel}}, \ and\ \bibinfo {author}
  {\bibfnamefont {L.}~\bibnamefont {Xu}},\ }\href@noop {} {\bibfield  {journal}
  {\bibinfo  {journal} {Chemical reviews}\ }\textbf {\bibinfo {volume} {116}},\
  \bibinfo {pages} {7608} (\bibinfo {year} {2016})}\BibitemShut {NoStop}%
\bibitem [{\citenamefont {Farnam}\ \emph
  {et~al.}(2014{\natexlab{a}})\citenamefont {Farnam}, \citenamefont {Bentz},
  \citenamefont {Sakulich}, \citenamefont {Flynn},\ and\ \citenamefont
  {Weiss}}]{farnam2014measuring}%
  \BibitemOpen
  \bibfield  {author} {\bibinfo {author} {\bibfnamefont {Y.}~\bibnamefont
  {Farnam}}, \bibinfo {author} {\bibfnamefont {D.}~\bibnamefont {Bentz}},
  \bibinfo {author} {\bibfnamefont {A.}~\bibnamefont {Sakulich}}, \bibinfo
  {author} {\bibfnamefont {D.}~\bibnamefont {Flynn}}, \ and\ \bibinfo {author}
  {\bibfnamefont {J.}~\bibnamefont {Weiss}},\ }\href@noop {} {\bibfield
  {journal} {\bibinfo  {journal} {Advances in Civil Engineering Materials}\
  }\textbf {\bibinfo {volume} {3}},\ \bibinfo {pages} {316} (\bibinfo {year}
  {2014}{\natexlab{a}})}\BibitemShut {NoStop}%
\bibitem [{\citenamefont {Cai}\ and\ \citenamefont
  {Liu}(1998)}]{cai1998freeze}%
  \BibitemOpen
  \bibfield  {author} {\bibinfo {author} {\bibfnamefont {H.}~\bibnamefont
  {Cai}}\ and\ \bibinfo {author} {\bibfnamefont {X.}~\bibnamefont {Liu}},\
  }\href@noop {} {\bibfield  {journal} {\bibinfo  {journal} {Cement and
  Concrete Research}\ }\textbf {\bibinfo {volume} {28}},\ \bibinfo {pages}
  {1281} (\bibinfo {year} {1998})}\BibitemShut {NoStop}%
\bibitem [{\citenamefont {Association}(2018)}]{PCAwebsitefreezethaw}%
  \BibitemOpen
  \bibfield  {author} {\bibinfo {author} {\bibfnamefont {P.~C.}\ \bibnamefont
  {Association}},\ }\href
  {https://www.cement.org/Learn/concrete-technology/durability/freeze-thaw-resistance}
  {\enquote {\bibinfo {title} {Freeze-thaw resistance},}\ } (\bibinfo {year}
  {2018})\BibitemShut {NoStop}%
\bibitem [{\citenamefont {Beaudoin}\ and\ \citenamefont
  {MacInnis}(1974)}]{beaudoin1974mechanism}%
  \BibitemOpen
  \bibfield  {author} {\bibinfo {author} {\bibfnamefont {J.~J.}\ \bibnamefont
  {Beaudoin}}\ and\ \bibinfo {author} {\bibfnamefont {C.}~\bibnamefont
  {MacInnis}},\ }\href@noop {} {\bibfield  {journal} {\bibinfo  {journal}
  {Cement and Concrete Research}\ }\textbf {\bibinfo {volume} {4}},\ \bibinfo
  {pages} {139} (\bibinfo {year} {1974})}\BibitemShut {NoStop}%
\bibitem [{\citenamefont {Winkler}\ and\ \citenamefont
  {Singer}(1972)}]{winkler1972crystallization}%
  \BibitemOpen
  \bibfield  {author} {\bibinfo {author} {\bibfnamefont {E.~M.}\ \bibnamefont
  {Winkler}}\ and\ \bibinfo {author} {\bibfnamefont {P.~C.}\ \bibnamefont
  {Singer}},\ }\href@noop {} {\bibfield  {journal} {\bibinfo  {journal}
  {Geological society of America bulletin}\ }\textbf {\bibinfo {volume} {83}},\
  \bibinfo {pages} {3509} (\bibinfo {year} {1972})}\BibitemShut {NoStop}%
\bibitem [{\citenamefont {Steiger}(2005)}]{steiger2005crystal}%
  \BibitemOpen
  \bibfield  {author} {\bibinfo {author} {\bibfnamefont {M.}~\bibnamefont
  {Steiger}},\ }\href@noop {} {\bibfield  {journal} {\bibinfo  {journal}
  {Journal of crystal growth}\ }\textbf {\bibinfo {volume} {282}},\ \bibinfo
  {pages} {455} (\bibinfo {year} {2005})}\BibitemShut {NoStop}%
\bibitem [{\citenamefont {Farnam}\ \emph
  {et~al.}(2014{\natexlab{b}})\citenamefont {Farnam}, \citenamefont {Bentz},
  \citenamefont {Hampton},\ and\ \citenamefont {Weiss}}]{farnam2014acoustic}%
  \BibitemOpen
  \bibfield  {author} {\bibinfo {author} {\bibfnamefont {Y.}~\bibnamefont
  {Farnam}}, \bibinfo {author} {\bibfnamefont {D.}~\bibnamefont {Bentz}},
  \bibinfo {author} {\bibfnamefont {A.}~\bibnamefont {Hampton}}, \ and\
  \bibinfo {author} {\bibfnamefont {W.}~\bibnamefont {Weiss}},\ }\href@noop {}
  {\bibfield  {journal} {\bibinfo  {journal} {Transportation Research Record:
  Journal of the Transportation Research Board}\ ,\ \bibinfo {pages} {81}}
  (\bibinfo {year} {2014}{\natexlab{b}})}\BibitemShut {NoStop}%
\bibitem [{\citenamefont {Li}\ \emph {et~al.}(2011)\citenamefont {Li},
  \citenamefont {Pour-Ghaz}, \citenamefont {Castro},\ and\ \citenamefont
  {Weiss}}]{li2011water}%
  \BibitemOpen
  \bibfield  {author} {\bibinfo {author} {\bibfnamefont {W.}~\bibnamefont
  {Li}}, \bibinfo {author} {\bibfnamefont {M.}~\bibnamefont {Pour-Ghaz}},
  \bibinfo {author} {\bibfnamefont {J.}~\bibnamefont {Castro}}, \ and\ \bibinfo
  {author} {\bibfnamefont {J.}~\bibnamefont {Weiss}},\ }\href@noop {}
  {\bibfield  {journal} {\bibinfo  {journal} {Journal of Materials in Civil
  Engineering}\ }\textbf {\bibinfo {volume} {24}},\ \bibinfo {pages} {299}
  (\bibinfo {year} {2011})}\BibitemShut {NoStop}%
\bibitem [{\citenamefont {Wettlaufer}\ and\ \citenamefont
  {Worster}(2006)}]{wettlaufer2006premelting}%
  \BibitemOpen
  \bibfield  {author} {\bibinfo {author} {\bibfnamefont {J.}~\bibnamefont
  {Wettlaufer}}\ and\ \bibinfo {author} {\bibfnamefont {M.~G.}\ \bibnamefont
  {Worster}},\ }\href@noop {} {\bibfield  {journal} {\bibinfo  {journal} {Annu.
  Rev. Fluid Mech.}\ }\textbf {\bibinfo {volume} {38}},\ \bibinfo {pages} {427}
  (\bibinfo {year} {2006})}\BibitemShut {NoStop}%
\bibitem [{\citenamefont {Zhou}\ \emph
  {et~al.}(2019{\natexlab{a}})\citenamefont {Zhou}, \citenamefont {Ioannidou},
  \citenamefont {Masoero}, \citenamefont {Mirzadeh}, \citenamefont {Pellenq},\
  and\ \citenamefont {Bazant}}]{zhou2019capillary}%
  \BibitemOpen
  \bibfield  {author} {\bibinfo {author} {\bibfnamefont {T.}~\bibnamefont
  {Zhou}}, \bibinfo {author} {\bibfnamefont {K.}~\bibnamefont {Ioannidou}},
  \bibinfo {author} {\bibfnamefont {E.}~\bibnamefont {Masoero}}, \bibinfo
  {author} {\bibfnamefont {M.}~\bibnamefont {Mirzadeh}}, \bibinfo {author}
  {\bibfnamefont {R.~J.-M.}\ \bibnamefont {Pellenq}}, \ and\ \bibinfo {author}
  {\bibfnamefont {M.~Z.}\ \bibnamefont {Bazant}},\ }\href@noop {} {\bibfield
  {journal} {\bibinfo  {journal} {Langmuir}\ }\textbf {\bibinfo {volume}
  {35}},\ \bibinfo {pages} {4397} (\bibinfo {year}
  {2019}{\natexlab{a}})}\BibitemShut {NoStop}%
\bibitem [{\citenamefont {Zhou}\ \emph
  {et~al.}(2019{\natexlab{b}})\citenamefont {Zhou}, \citenamefont {Ioannidou},
  \citenamefont {Ulm}, \citenamefont {Bazant},\ and\ \citenamefont
  {Pellenq}}]{zhou2019multiscale}%
  \BibitemOpen
  \bibfield  {author} {\bibinfo {author} {\bibfnamefont {T.}~\bibnamefont
  {Zhou}}, \bibinfo {author} {\bibfnamefont {K.}~\bibnamefont {Ioannidou}},
  \bibinfo {author} {\bibfnamefont {F.-J.}\ \bibnamefont {Ulm}}, \bibinfo
  {author} {\bibfnamefont {M.~Z.}\ \bibnamefont {Bazant}}, \ and\ \bibinfo
  {author} {\bibfnamefont {R.-M.}\ \bibnamefont {Pellenq}},\ }\href@noop {}
  {\bibfield  {journal} {\bibinfo  {journal} {Proceedings of the National
  Academy of Sciences}\ }\textbf {\bibinfo {volume} {116}},\ \bibinfo {pages}
  {10652} (\bibinfo {year} {2019}{\natexlab{b}})}\BibitemShut {NoStop}%
\bibitem [{\citenamefont {Lyklema}(1995)}]{lyklema_book_vol2}%
  \BibitemOpen
  \bibfield  {author} {\bibinfo {author} {\bibfnamefont {J.}~\bibnamefont
  {Lyklema}},\ }\href@noop {} {\emph {\bibinfo {title} {Fundamentals of
  Interface and Colloid Science. Volume II: Solid-Liquid Interfaces}}}\
  (\bibinfo  {publisher} {Academic Press Limited},\ \bibinfo {address} {San
  Diego, CA},\ \bibinfo {year} {1995})\BibitemShut {NoStop}%
\bibitem [{\citenamefont {Israelachvili}(2011)}]{israelachvili_book}%
  \BibitemOpen
  \bibfield  {author} {\bibinfo {author} {\bibfnamefont {J.~N.}\ \bibnamefont
  {Israelachvili}},\ }\href@noop {} {\emph {\bibinfo {title} {Intermolecular
  and Surface Forces}}}\ (\bibinfo  {publisher} {Academic Press},\ \bibinfo
  {address} {New York, NY},\ \bibinfo {year} {2011})\BibitemShut {NoStop}%
\bibitem [{\citenamefont {Andelman}(1995)}]{andelman1995_PBforces}%
  \BibitemOpen
  \bibfield  {author} {\bibinfo {author} {\bibfnamefont {D.}~\bibnamefont
  {Andelman}},\ }in\ \href@noop {} {\emph {\bibinfo {booktitle} {Handbook of
  Biological Physics}}}\ (\bibinfo  {publisher} {Elsevier},\ \bibinfo {year}
  {1995})\ pp.\ \bibinfo {pages} {603--641}\BibitemShut {NoStop}%
\bibitem [{\citenamefont {Pellenq}\ and\ \citenamefont
  {Van~Damme}(2004)}]{pellenq2004does}%
  \BibitemOpen
  \bibfield  {author} {\bibinfo {author} {\bibfnamefont {R.~J.-M.}\
  \bibnamefont {Pellenq}}\ and\ \bibinfo {author} {\bibfnamefont
  {H.}~\bibnamefont {Van~Damme}},\ }\href@noop {} {\bibfield  {journal}
  {\bibinfo  {journal} {Mrs Bulletin}\ }\textbf {\bibinfo {volume} {29}},\
  \bibinfo {pages} {319} (\bibinfo {year} {2004})}\BibitemShut {NoStop}%
\bibitem [{\citenamefont {Pellenq}\ \emph {et~al.}(1997)\citenamefont
  {Pellenq}, \citenamefont {Caillol},\ and\ \citenamefont
  {Delville}}]{pellenq1997electrostatic}%
  \BibitemOpen
  \bibfield  {author} {\bibinfo {author} {\bibfnamefont {R.-M.}\ \bibnamefont
  {Pellenq}}, \bibinfo {author} {\bibfnamefont {J.}~\bibnamefont {Caillol}}, \
  and\ \bibinfo {author} {\bibfnamefont {A.}~\bibnamefont {Delville}},\
  }\href@noop {} {\bibfield  {journal} {\bibinfo  {journal} {The Journal of
  Physical Chemistry B}\ }\textbf {\bibinfo {volume} {101}},\ \bibinfo {pages}
  {8584} (\bibinfo {year} {1997})}\BibitemShut {NoStop}%
\bibitem [{\citenamefont {Alba-Simionesco}\ \emph {et~al.}(2006)\citenamefont
  {Alba-Simionesco}, \citenamefont {Coasne}, \citenamefont {Dosseh},
  \citenamefont {Dudziak}, \citenamefont {Gubbins}, \citenamefont
  {Radhakrishnan},\ and\ \citenamefont
  {Sliwinska-Bartkowiak}}]{alba2006effects}%
  \BibitemOpen
  \bibfield  {author} {\bibinfo {author} {\bibfnamefont {C.}~\bibnamefont
  {Alba-Simionesco}}, \bibinfo {author} {\bibfnamefont {B.}~\bibnamefont
  {Coasne}}, \bibinfo {author} {\bibfnamefont {G.}~\bibnamefont {Dosseh}},
  \bibinfo {author} {\bibfnamefont {G.}~\bibnamefont {Dudziak}}, \bibinfo
  {author} {\bibfnamefont {K.}~\bibnamefont {Gubbins}}, \bibinfo {author}
  {\bibfnamefont {R.}~\bibnamefont {Radhakrishnan}}, \ and\ \bibinfo {author}
  {\bibfnamefont {M.}~\bibnamefont {Sliwinska-Bartkowiak}},\ }\href@noop {}
  {\bibfield  {journal} {\bibinfo  {journal} {Journal of Physics: Condensed
  Matter}\ }\textbf {\bibinfo {volume} {18}},\ \bibinfo {pages} {R15} (\bibinfo
  {year} {2006})}\BibitemShut {NoStop}%
\bibitem [{\citenamefont {Jan}\ and\ \citenamefont
  {Chien}(1973)}]{jan1973role}%
  \BibitemOpen
  \bibfield  {author} {\bibinfo {author} {\bibfnamefont {K.-M.}\ \bibnamefont
  {Jan}}\ and\ \bibinfo {author} {\bibfnamefont {S.}~\bibnamefont {Chien}},\
  }\href@noop {} {\bibfield  {journal} {\bibinfo  {journal} {The Journal of
  general physiology}\ }\textbf {\bibinfo {volume} {61}},\ \bibinfo {pages}
  {638} (\bibinfo {year} {1973})}\BibitemShut {NoStop}%
\bibitem [{\citenamefont {Visser}(1973)}]{visser1973adhesion}%
  \BibitemOpen
  \bibfield  {author} {\bibinfo {author} {\bibfnamefont {J.}~\bibnamefont
  {Visser}},\ }\href@noop {} {\  (\bibinfo {year} {1973})}\BibitemShut
  {NoStop}%
\bibitem [{\citenamefont {Giese}(1973)}]{giese1973interlayer}%
  \BibitemOpen
  \bibfield  {author} {\bibinfo {author} {\bibfnamefont {R.}~\bibnamefont
  {Giese}},\ }\href@noop {} {\bibfield  {journal} {\bibinfo  {journal} {Clays
  and Clay Minerals}\ }\textbf {\bibinfo {volume} {21}},\ \bibinfo {pages}
  {145} (\bibinfo {year} {1973})}\BibitemShut {NoStop}%
\bibitem [{\citenamefont {McLaughlin}(1977)}]{mclaughlin1977electrostatic}%
  \BibitemOpen
  \bibfield  {author} {\bibinfo {author} {\bibfnamefont {S.}~\bibnamefont
  {McLaughlin}},\ }in\ \href@noop {} {\emph {\bibinfo {booktitle} {Current
  topics in membranes and transport}}},\ Vol.~\bibinfo {volume} {9}\ (\bibinfo
  {publisher} {Elsevier},\ \bibinfo {year} {1977})\ pp.\ \bibinfo {pages}
  {71--144}\BibitemShut {NoStop}%
\bibitem [{\citenamefont {Swartzen-Allen}\ and\ \citenamefont
  {Matijevic}(1974)}]{swartzen1974surface}%
  \BibitemOpen
  \bibfield  {author} {\bibinfo {author} {\bibfnamefont {S.~L.}\ \bibnamefont
  {Swartzen-Allen}}\ and\ \bibinfo {author} {\bibfnamefont {E.}~\bibnamefont
  {Matijevic}},\ }\href@noop {} {\bibfield  {journal} {\bibinfo  {journal}
  {Chemical Reviews}\ }\textbf {\bibinfo {volume} {74}},\ \bibinfo {pages}
  {385} (\bibinfo {year} {1974})}\BibitemShut {NoStop}%
\bibitem [{\citenamefont {Allen}\ \emph {et~al.}(2007)\citenamefont {Allen},
  \citenamefont {Thomas},\ and\ \citenamefont
  {Jennings}}]{allen2007composition}%
  \BibitemOpen
  \bibfield  {author} {\bibinfo {author} {\bibfnamefont {A.~J.}\ \bibnamefont
  {Allen}}, \bibinfo {author} {\bibfnamefont {J.~J.}\ \bibnamefont {Thomas}}, \
  and\ \bibinfo {author} {\bibfnamefont {H.~M.}\ \bibnamefont {Jennings}},\
  }\href@noop {} {\bibfield  {journal} {\bibinfo  {journal} {Nature materials}\
  }\textbf {\bibinfo {volume} {6}},\ \bibinfo {pages} {311} (\bibinfo {year}
  {2007})}\BibitemShut {NoStop}%
\bibitem [{\citenamefont {Pellenq}\ \emph {et~al.}(2009)\citenamefont
  {Pellenq}, \citenamefont {Kushima}, \citenamefont {Shahsavari}, \citenamefont
  {Van~Vliet}, \citenamefont {Buehler}, \citenamefont {Yip},\ and\
  \citenamefont {Ulm}}]{pellenq2009realistic}%
  \BibitemOpen
  \bibfield  {author} {\bibinfo {author} {\bibfnamefont {R.~J.-M.}\
  \bibnamefont {Pellenq}}, \bibinfo {author} {\bibfnamefont {A.}~\bibnamefont
  {Kushima}}, \bibinfo {author} {\bibfnamefont {R.}~\bibnamefont {Shahsavari}},
  \bibinfo {author} {\bibfnamefont {K.~J.}\ \bibnamefont {Van~Vliet}}, \bibinfo
  {author} {\bibfnamefont {M.~J.}\ \bibnamefont {Buehler}}, \bibinfo {author}
  {\bibfnamefont {S.}~\bibnamefont {Yip}}, \ and\ \bibinfo {author}
  {\bibfnamefont {F.-J.}\ \bibnamefont {Ulm}},\ }\href@noop {} {\bibfield
  {journal} {\bibinfo  {journal} {Proceedings of the National Academy of
  Sciences}\ }\textbf {\bibinfo {volume} {106}},\ \bibinfo {pages} {16102}
  (\bibinfo {year} {2009})}\BibitemShut {NoStop}%
\bibitem [{\citenamefont {Delville}\ \emph {et~al.}(1998)\citenamefont
  {Delville}, \citenamefont {Gasmi}, \citenamefont {Pellenq}, \citenamefont
  {Caillol},\ and\ \citenamefont {Van~Damme}}]{delville1998correlations}%
  \BibitemOpen
  \bibfield  {author} {\bibinfo {author} {\bibfnamefont {A.}~\bibnamefont
  {Delville}}, \bibinfo {author} {\bibfnamefont {N.}~\bibnamefont {Gasmi}},
  \bibinfo {author} {\bibfnamefont {R.~J.}\ \bibnamefont {Pellenq}}, \bibinfo
  {author} {\bibfnamefont {J.~M.}\ \bibnamefont {Caillol}}, \ and\ \bibinfo
  {author} {\bibfnamefont {H.}~\bibnamefont {Van~Damme}},\ }\href@noop {}
  {\bibfield  {journal} {\bibinfo  {journal} {Langmuir}\ }\textbf {\bibinfo
  {volume} {14}},\ \bibinfo {pages} {5077} (\bibinfo {year}
  {1998})}\BibitemShut {NoStop}%
\bibitem [{\citenamefont {Jellander}\ \emph {et~al.}(1988)\citenamefont
  {Jellander}, \citenamefont {Mar{\v{c}}elja},\ and\ \citenamefont
  {Quirk}}]{jellander1988attractive}%
  \BibitemOpen
  \bibfield  {author} {\bibinfo {author} {\bibfnamefont {R.}~\bibnamefont
  {Jellander}}, \bibinfo {author} {\bibfnamefont {S.}~\bibnamefont
  {Mar{\v{c}}elja}}, \ and\ \bibinfo {author} {\bibfnamefont {J.}~\bibnamefont
  {Quirk}},\ }\href@noop {} {\bibfield  {journal} {\bibinfo  {journal} {Journal
  of Colloid and Interface Science}\ }\textbf {\bibinfo {volume} {126}},\
  \bibinfo {pages} {194} (\bibinfo {year} {1988})}\BibitemShut {NoStop}%
\bibitem [{\citenamefont {Stankovich}\ and\ \citenamefont
  {Carnie}(1996)}]{stankovich1996electrical}%
  \BibitemOpen
  \bibfield  {author} {\bibinfo {author} {\bibfnamefont {J.}~\bibnamefont
  {Stankovich}}\ and\ \bibinfo {author} {\bibfnamefont {S.~L.}\ \bibnamefont
  {Carnie}},\ }\href@noop {} {\bibfield  {journal} {\bibinfo  {journal}
  {Langmuir}\ }\textbf {\bibinfo {volume} {12}},\ \bibinfo {pages} {1453}
  (\bibinfo {year} {1996})}\BibitemShut {NoStop}%
\bibitem [{\citenamefont {Valleau}\ and\ \citenamefont
  {Cohen}(1980)}]{valleau1980primitive}%
  \BibitemOpen
  \bibfield  {author} {\bibinfo {author} {\bibfnamefont {J.~P.}\ \bibnamefont
  {Valleau}}\ and\ \bibinfo {author} {\bibfnamefont {L.~K.}\ \bibnamefont
  {Cohen}},\ }\href@noop {} {\bibfield  {journal} {\bibinfo  {journal} {The
  Journal of chemical physics}\ }\textbf {\bibinfo {volume} {72}},\ \bibinfo
  {pages} {5935} (\bibinfo {year} {1980})}\BibitemShut {NoStop}%
\bibitem [{\citenamefont {Torquato}(2013)}]{torquato2013random}%
  \BibitemOpen
  \bibfield  {author} {\bibinfo {author} {\bibfnamefont {S.}~\bibnamefont
  {Torquato}},\ }\href@noop {} {\emph {\bibinfo {title} {Random heterogeneous
  materials: microstructure and macroscopic properties}}},\ Vol.~\bibinfo
  {volume} {16}\ (\bibinfo  {publisher} {Springer Science \& Business Media},\
  \bibinfo {year} {2013})\BibitemShut {NoStop}%
\bibitem [{\citenamefont {Van~Brakel}(1975)}]{van1975pore}%
  \BibitemOpen
  \bibfield  {author} {\bibinfo {author} {\bibfnamefont {J.}~\bibnamefont
  {Van~Brakel}},\ }\href@noop {} {\bibfield  {journal} {\bibinfo  {journal}
  {Powder technology}\ }\textbf {\bibinfo {volume} {11}},\ \bibinfo {pages}
  {205} (\bibinfo {year} {1975})}\BibitemShut {NoStop}%
\bibitem [{\citenamefont {Quiblier}(1984)}]{quiblier1984new}%
  \BibitemOpen
  \bibfield  {author} {\bibinfo {author} {\bibfnamefont {J.~A.}\ \bibnamefont
  {Quiblier}},\ }\href@noop {} {\bibfield  {journal} {\bibinfo  {journal}
  {Journal of Colloid and Interface Science}\ }\textbf {\bibinfo {volume}
  {98}},\ \bibinfo {pages} {84} (\bibinfo {year} {1984})}\BibitemShut {NoStop}%
\bibitem [{\citenamefont {Pinson}\ \emph {et~al.}(2018)\citenamefont {Pinson},
  \citenamefont {Zhou}, \citenamefont {Jennings},\ and\ \citenamefont
  {Bazant}}]{pinson2018inferring}%
  \BibitemOpen
  \bibfield  {author} {\bibinfo {author} {\bibfnamefont {M.~B.}\ \bibnamefont
  {Pinson}}, \bibinfo {author} {\bibfnamefont {T.}~\bibnamefont {Zhou}},
  \bibinfo {author} {\bibfnamefont {H.~M.}\ \bibnamefont {Jennings}}, \ and\
  \bibinfo {author} {\bibfnamefont {M.~Z.}\ \bibnamefont {Bazant}},\
  }\href@noop {} {\bibfield  {journal} {\bibinfo  {journal} {Journal of colloid
  and interface science}\ } (\bibinfo {year} {2018})}\BibitemShut {NoStop}%
\bibitem [{\citenamefont {Gu}\ and\ \citenamefont
  {Bazant}(2018)}]{gu2018microscopic}%
  \BibitemOpen
  \bibfield  {author} {\bibinfo {author} {\bibfnamefont {Z.}~\bibnamefont
  {Gu}}\ and\ \bibinfo {author} {\bibfnamefont {M.~Z.}\ \bibnamefont
  {Bazant}},\ }\href@noop {} {\bibfield  {journal} {\bibinfo  {journal} {arXiv
  preprint arXiv:1808.09804}\ } (\bibinfo {year} {2018})}\BibitemShut {NoStop}%
\bibitem [{\citenamefont {Biesheuvel}\ and\ \citenamefont
  {Bazant}(2016)}]{biesheuvel2016analysis}%
  \BibitemOpen
  \bibfield  {author} {\bibinfo {author} {\bibfnamefont {P.}~\bibnamefont
  {Biesheuvel}}\ and\ \bibinfo {author} {\bibfnamefont {M.}~\bibnamefont
  {Bazant}},\ }\href@noop {} {\bibfield  {journal} {\bibinfo  {journal}
  {Physical Review E}\ }\textbf {\bibinfo {volume} {94}},\ \bibinfo {pages}
  {050601} (\bibinfo {year} {2016})}\BibitemShut {NoStop}%
\bibitem [{\citenamefont {Peters}\ \emph {et~al.}(2016)\citenamefont {Peters},
  \citenamefont {Van~Roij}, \citenamefont {Bazant},\ and\ \citenamefont
  {Biesheuvel}}]{peters2016analysis}%
  \BibitemOpen
  \bibfield  {author} {\bibinfo {author} {\bibfnamefont {P.}~\bibnamefont
  {Peters}}, \bibinfo {author} {\bibfnamefont {R.}~\bibnamefont {Van~Roij}},
  \bibinfo {author} {\bibfnamefont {M.~Z.}\ \bibnamefont {Bazant}}, \ and\
  \bibinfo {author} {\bibfnamefont {P.}~\bibnamefont {Biesheuvel}},\
  }\href@noop {} {\bibfield  {journal} {\bibinfo  {journal} {Physical review
  E}\ }\textbf {\bibinfo {volume} {93}},\ \bibinfo {pages} {053108} (\bibinfo
  {year} {2016})}\BibitemShut {NoStop}%
\bibitem [{\citenamefont {Catalano}\ \emph {et~al.}(2016)\citenamefont
  {Catalano}, \citenamefont {Lammertink},\ and\ \citenamefont
  {Biesheuvel}}]{catalano2016theory}%
  \BibitemOpen
  \bibfield  {author} {\bibinfo {author} {\bibfnamefont {J.}~\bibnamefont
  {Catalano}}, \bibinfo {author} {\bibfnamefont {R.}~\bibnamefont
  {Lammertink}}, \ and\ \bibinfo {author} {\bibfnamefont {P.}~\bibnamefont
  {Biesheuvel}},\ }\href@noop {} {\bibfield  {journal} {\bibinfo  {journal}
  {arXiv preprint arXiv:1603.09293}\ } (\bibinfo {year} {2016})}\BibitemShut
  {NoStop}%
\bibitem [{\citenamefont {Yossifon}\ \emph {et~al.}(2006)\citenamefont
  {Yossifon}, \citenamefont {Frankel},\ and\ \citenamefont
  {Miloh}}]{yossifon2006electro}%
  \BibitemOpen
  \bibfield  {author} {\bibinfo {author} {\bibfnamefont {G.}~\bibnamefont
  {Yossifon}}, \bibinfo {author} {\bibfnamefont {I.}~\bibnamefont {Frankel}}, \
  and\ \bibinfo {author} {\bibfnamefont {T.}~\bibnamefont {Miloh}},\
  }\href@noop {} {\bibfield  {journal} {\bibinfo  {journal} {Physics of
  Fluids}\ }\textbf {\bibinfo {volume} {18}},\ \bibinfo {pages} {117108}
  (\bibinfo {year} {2006})}\BibitemShut {NoStop}%
\bibitem [{\citenamefont {Pennathur}\ and\ \citenamefont
  {Santiago}(2005)}]{pennathur2005electrokinetic}%
  \BibitemOpen
  \bibfield  {author} {\bibinfo {author} {\bibfnamefont {S.}~\bibnamefont
  {Pennathur}}\ and\ \bibinfo {author} {\bibfnamefont {J.~G.}\ \bibnamefont
  {Santiago}},\ }\href@noop {} {\bibfield  {journal} {\bibinfo  {journal}
  {Analytical chemistry}\ }\textbf {\bibinfo {volume} {77}},\ \bibinfo {pages}
  {6772} (\bibinfo {year} {2005})}\BibitemShut {NoStop}%
\bibitem [{\citenamefont {Chang}\ \emph {et~al.}(2012)\citenamefont {Chang},
  \citenamefont {Yossifon},\ and\ \citenamefont
  {Demekhin}}]{chang2012nanoscale}%
  \BibitemOpen
  \bibfield  {author} {\bibinfo {author} {\bibfnamefont {H.-C.}\ \bibnamefont
  {Chang}}, \bibinfo {author} {\bibfnamefont {G.}~\bibnamefont {Yossifon}}, \
  and\ \bibinfo {author} {\bibfnamefont {E.~A.}\ \bibnamefont {Demekhin}},\
  }\href@noop {} {\bibfield  {journal} {\bibinfo  {journal} {Annual review of
  fluid mechanics}\ }\textbf {\bibinfo {volume} {44}},\ \bibinfo {pages} {401}
  (\bibinfo {year} {2012})}\BibitemShut {NoStop}%
\bibitem [{\citenamefont {Yossifon}\ \emph
  {et~al.}(2009{\natexlab{a}})\citenamefont {Yossifon}, \citenamefont
  {Mushenheim}, \citenamefont {Chang},\ and\ \citenamefont
  {Chang}}]{yossifon2009nonlinear}%
  \BibitemOpen
  \bibfield  {author} {\bibinfo {author} {\bibfnamefont {G.}~\bibnamefont
  {Yossifon}}, \bibinfo {author} {\bibfnamefont {P.}~\bibnamefont
  {Mushenheim}}, \bibinfo {author} {\bibfnamefont {Y.-C.}\ \bibnamefont
  {Chang}}, \ and\ \bibinfo {author} {\bibfnamefont {H.-C.}\ \bibnamefont
  {Chang}},\ }\href@noop {} {\bibfield  {journal} {\bibinfo  {journal}
  {Physical Review E}\ }\textbf {\bibinfo {volume} {79}},\ \bibinfo {pages}
  {046305} (\bibinfo {year} {2009}{\natexlab{a}})}\BibitemShut {NoStop}%
\bibitem [{\citenamefont {Denoyel}\ and\ \citenamefont
  {Pellenq}(2002)}]{denoyel2002simple}%
  \BibitemOpen
  \bibfield  {author} {\bibinfo {author} {\bibfnamefont {R.}~\bibnamefont
  {Denoyel}}\ and\ \bibinfo {author} {\bibfnamefont {R.}~\bibnamefont
  {Pellenq}},\ }\href@noop {} {\bibfield  {journal} {\bibinfo  {journal}
  {Langmuir}\ }\textbf {\bibinfo {volume} {18}},\ \bibinfo {pages} {2710}
  (\bibinfo {year} {2002})}\BibitemShut {NoStop}%
\bibitem [{\citenamefont {Bertil}(1992)}]{hille2001ion}%
  \BibitemOpen
  \bibfield  {author} {\bibinfo {author} {\bibfnamefont {H.}~\bibnamefont
  {Bertil}},\ }\href@noop {} {\emph {\bibinfo {title} {Ionic channels of
  excitable membranes}}}\ (\bibinfo  {publisher} {Sinauer Sunderland, MA},\
  \bibinfo {year} {1992})\BibitemShut {NoStop}%
\bibitem [{\citenamefont {Born}\ and\ \citenamefont
  {Oppenheimer}(1927)}]{born1927quantentheorie}%
  \BibitemOpen
  \bibfield  {author} {\bibinfo {author} {\bibfnamefont {M.}~\bibnamefont
  {Born}}\ and\ \bibinfo {author} {\bibfnamefont {R.}~\bibnamefont
  {Oppenheimer}},\ }\href@noop {} {\bibfield  {journal} {\bibinfo  {journal}
  {Annalen der physik}\ }\textbf {\bibinfo {volume} {389}},\ \bibinfo {pages}
  {457} (\bibinfo {year} {1927})}\BibitemShut {NoStop}%
\bibitem [{\citenamefont {Zeng}\ \emph {et~al.}(2001)\citenamefont {Zeng},
  \citenamefont {Chen}, \citenamefont {Mikkelsen~Jr},\ and\ \citenamefont
  {Santiago}}]{zeng2001fabrication}%
  \BibitemOpen
  \bibfield  {author} {\bibinfo {author} {\bibfnamefont {S.}~\bibnamefont
  {Zeng}}, \bibinfo {author} {\bibfnamefont {C.-H.}\ \bibnamefont {Chen}},
  \bibinfo {author} {\bibfnamefont {J.~C.}\ \bibnamefont {Mikkelsen~Jr}}, \
  and\ \bibinfo {author} {\bibfnamefont {J.~G.}\ \bibnamefont {Santiago}},\
  }\href@noop {} {\bibfield  {journal} {\bibinfo  {journal} {Sensors and
  Actuators B: Chemical}\ }\textbf {\bibinfo {volume} {79}},\ \bibinfo {pages}
  {107} (\bibinfo {year} {2001})}\BibitemShut {NoStop}%
\bibitem [{\citenamefont {Daiguji}\ \emph {et~al.}(2005)\citenamefont
  {Daiguji}, \citenamefont {Oka},\ and\ \citenamefont
  {Shirono}}]{daiguji2005nanofluidic}%
  \BibitemOpen
  \bibfield  {author} {\bibinfo {author} {\bibfnamefont {H.}~\bibnamefont
  {Daiguji}}, \bibinfo {author} {\bibfnamefont {Y.}~\bibnamefont {Oka}}, \ and\
  \bibinfo {author} {\bibfnamefont {K.}~\bibnamefont {Shirono}},\ }\href@noop
  {} {\bibfield  {journal} {\bibinfo  {journal} {Nano Letters}\ }\textbf
  {\bibinfo {volume} {5}},\ \bibinfo {pages} {2274} (\bibinfo {year}
  {2005})}\BibitemShut {NoStop}%
\bibitem [{\citenamefont {Yossifon}\ \emph
  {et~al.}(2009{\natexlab{b}})\citenamefont {Yossifon}, \citenamefont {Chang},\
  and\ \citenamefont {Chang}}]{yossifon2009rectification}%
  \BibitemOpen
  \bibfield  {author} {\bibinfo {author} {\bibfnamefont {G.}~\bibnamefont
  {Yossifon}}, \bibinfo {author} {\bibfnamefont {Y.-C.}\ \bibnamefont {Chang}},
  \ and\ \bibinfo {author} {\bibfnamefont {H.-C.}\ \bibnamefont {Chang}},\
  }\href@noop {} {\bibfield  {journal} {\bibinfo  {journal} {Physical review
  letters}\ }\textbf {\bibinfo {volume} {103}},\ \bibinfo {pages} {154502}
  (\bibinfo {year} {2009}{\natexlab{b}})}\BibitemShut {NoStop}%
\bibitem [{\citenamefont {Gillespie}\ and\ \citenamefont
  {Pennathur}(2013)}]{gillespie2013separation}%
  \BibitemOpen
  \bibfield  {author} {\bibinfo {author} {\bibfnamefont {D.}~\bibnamefont
  {Gillespie}}\ and\ \bibinfo {author} {\bibfnamefont {S.}~\bibnamefont
  {Pennathur}},\ }\href@noop {} {\bibfield  {journal} {\bibinfo  {journal}
  {Analytical chemistry}\ }\textbf {\bibinfo {volume} {85}},\ \bibinfo {pages}
  {2991} (\bibinfo {year} {2013})}\BibitemShut {NoStop}%
\bibitem [{\citenamefont {Bazant}\ \emph {et~al.}(2011)\citenamefont {Bazant},
  \citenamefont {Storey},\ and\ \citenamefont {Kornyshev}}]{bazant2011double}%
  \BibitemOpen
  \bibfield  {author} {\bibinfo {author} {\bibfnamefont {M.~Z.}\ \bibnamefont
  {Bazant}}, \bibinfo {author} {\bibfnamefont {B.~D.}\ \bibnamefont {Storey}},
  \ and\ \bibinfo {author} {\bibfnamefont {A.~A.}\ \bibnamefont {Kornyshev}},\
  }\href@noop {} {\bibfield  {journal} {\bibinfo  {journal} {Physical Review
  Letters}\ }\textbf {\bibinfo {volume} {106}},\ \bibinfo {pages} {046102}
  (\bibinfo {year} {2011})}\BibitemShut {NoStop}%
\bibitem [{\citenamefont {Bazant}\ \emph {et~al.}(2009)\citenamefont {Bazant},
  \citenamefont {Kilic}, \citenamefont {Storey},\ and\ \citenamefont
  {Ajdari}}]{bazant2009towards}%
  \BibitemOpen
  \bibfield  {author} {\bibinfo {author} {\bibfnamefont {M.~Z.}\ \bibnamefont
  {Bazant}}, \bibinfo {author} {\bibfnamefont {M.~S.}\ \bibnamefont {Kilic}},
  \bibinfo {author} {\bibfnamefont {B.}~\bibnamefont {Storey}}, \ and\ \bibinfo
  {author} {\bibfnamefont {A.}~\bibnamefont {Ajdari}},\ }\href@noop {}
  {\bibfield  {journal} {\bibinfo  {journal} {Advances in Colloid and Interface
  Science}\ }\textbf {\bibinfo {volume} {152}},\ \bibinfo {pages} {48}
  (\bibinfo {year} {2009})}\BibitemShut {NoStop}%
\bibitem [{\citenamefont {Bohinc}\ \emph {et~al.}(2012)\citenamefont {Bohinc},
  \citenamefont {Shrestha}, \citenamefont {Brumen},\ and\ \citenamefont
  {May}}]{bohinc2012poisson}%
  \BibitemOpen
  \bibfield  {author} {\bibinfo {author} {\bibfnamefont {K.}~\bibnamefont
  {Bohinc}}, \bibinfo {author} {\bibfnamefont {A.}~\bibnamefont {Shrestha}},
  \bibinfo {author} {\bibfnamefont {M.}~\bibnamefont {Brumen}}, \ and\ \bibinfo
  {author} {\bibfnamefont {S.}~\bibnamefont {May}},\ }\href@noop {} {\bibfield
  {journal} {\bibinfo  {journal} {Physical Review E}\ }\textbf {\bibinfo
  {volume} {85}},\ \bibinfo {pages} {031130} (\bibinfo {year}
  {2012})}\BibitemShut {NoStop}%
\bibitem [{\citenamefont {Brown}\ \emph {et~al.}(2015)\citenamefont {Brown},
  \citenamefont {Bossa},\ and\ \citenamefont {May}}]{brown2015emergence}%
  \BibitemOpen
  \bibfield  {author} {\bibinfo {author} {\bibfnamefont {M.~A.}\ \bibnamefont
  {Brown}}, \bibinfo {author} {\bibfnamefont {G.~V.}\ \bibnamefont {Bossa}}, \
  and\ \bibinfo {author} {\bibfnamefont {S.}~\bibnamefont {May}},\ }\href@noop
  {} {\bibfield  {journal} {\bibinfo  {journal} {Langmuir}\ }\textbf {\bibinfo
  {volume} {31}},\ \bibinfo {pages} {11477} (\bibinfo {year}
  {2015})}\BibitemShut {NoStop}%
\bibitem [{\citenamefont {Booth}(1951)}]{booth1951dielectric}%
  \BibitemOpen
  \bibfield  {author} {\bibinfo {author} {\bibfnamefont {F.}~\bibnamefont
  {Booth}},\ }\href@noop {} {\bibfield  {journal} {\bibinfo  {journal} {The
  Journal of Chemical Physics}\ }\textbf {\bibinfo {volume} {19}},\ \bibinfo
  {pages} {391} (\bibinfo {year} {1951})}\BibitemShut {NoStop}%
\bibitem [{\citenamefont {Aguilella-Arzo}\ \emph {et~al.}(2009)\citenamefont
  {Aguilella-Arzo}, \citenamefont {Andrio}, \citenamefont {Aguilella},\ and\
  \citenamefont {Alcaraz}}]{aguilella2009dielectric}%
  \BibitemOpen
  \bibfield  {author} {\bibinfo {author} {\bibfnamefont {M.}~\bibnamefont
  {Aguilella-Arzo}}, \bibinfo {author} {\bibfnamefont {A.}~\bibnamefont
  {Andrio}}, \bibinfo {author} {\bibfnamefont {V.~M.}\ \bibnamefont
  {Aguilella}}, \ and\ \bibinfo {author} {\bibfnamefont {A.}~\bibnamefont
  {Alcaraz}},\ }\href@noop {} {\bibfield  {journal} {\bibinfo  {journal}
  {Physical Chemistry Chemical Physics}\ }\textbf {\bibinfo {volume} {11}},\
  \bibinfo {pages} {358} (\bibinfo {year} {2009})}\BibitemShut {NoStop}%
\bibitem [{\citenamefont {Zhou}\ \emph {et~al.}()\citenamefont {Zhou},
  \citenamefont {Mirzadeh}, \citenamefont {Dimitrios~Fraggedakis},\ and\
  \citenamefont {Bazant}}]{zhou2019freezingtheory}%
  \BibitemOpen
  \bibfield  {author} {\bibinfo {author} {\bibfnamefont {T.}~\bibnamefont
  {Zhou}}, \bibinfo {author} {\bibfnamefont {M.}~\bibnamefont {Mirzadeh}},
  \bibinfo {author} {\bibfnamefont {R.~J.-M.~P.}\ \bibnamefont
  {Dimitrios~Fraggedakis}}, \ and\ \bibinfo {author} {\bibfnamefont {M.~Z.}\
  \bibnamefont {Bazant}},\ }\href@noop {} {\enquote {\bibinfo {title} {Theory
  of freezing point depression in charged porous media},}\ }\bibinfo {note} {In
  preparation}\BibitemShut {NoStop}%
\bibitem [{\citenamefont {Gibbs}(1879)}]{gibbs1879equilibrium}%
  \BibitemOpen
  \bibfield  {author} {\bibinfo {author} {\bibfnamefont {J.~W.}\ \bibnamefont
  {Gibbs}},\ }\href@noop {} {\  (\bibinfo {year} {1879})}\BibitemShut {NoStop}%
\bibitem [{\citenamefont {Brush}\ \emph {et~al.}(1966)\citenamefont {Brush},
  \citenamefont {Sahlin},\ and\ \citenamefont {Teller}}]{brush1966monte}%
  \BibitemOpen
  \bibfield  {author} {\bibinfo {author} {\bibfnamefont {S.}~\bibnamefont
  {Brush}}, \bibinfo {author} {\bibfnamefont {H.}~\bibnamefont {Sahlin}}, \
  and\ \bibinfo {author} {\bibfnamefont {E.}~\bibnamefont {Teller}},\
  }\href@noop {} {\bibfield  {journal} {\bibinfo  {journal} {The Journal of
  Chemical Physics}\ }\textbf {\bibinfo {volume} {45}},\ \bibinfo {pages}
  {2102} (\bibinfo {year} {1966})}\BibitemShut {NoStop}%
\bibitem [{\citenamefont {Baus}\ and\ \citenamefont
  {Hansen}(1980)}]{baus1980statistical}%
  \BibitemOpen
  \bibfield  {author} {\bibinfo {author} {\bibfnamefont {M.}~\bibnamefont
  {Baus}}\ and\ \bibinfo {author} {\bibfnamefont {J.-P.}\ \bibnamefont
  {Hansen}},\ }\href@noop {} {\bibfield  {journal} {\bibinfo  {journal}
  {Physics Reports}\ }\textbf {\bibinfo {volume} {59}},\ \bibinfo {pages} {1}
  (\bibinfo {year} {1980})}\BibitemShut {NoStop}%
\bibitem [{\citenamefont {Markovich}\ \emph {et~al.}(2016)\citenamefont
  {Markovich}, \citenamefont {Andelman},\ and\ \citenamefont
  {Podgornik}}]{markovich2016charge}%
  \BibitemOpen
  \bibfield  {author} {\bibinfo {author} {\bibfnamefont {T.}~\bibnamefont
  {Markovich}}, \bibinfo {author} {\bibfnamefont {D.}~\bibnamefont {Andelman}},
  \ and\ \bibinfo {author} {\bibfnamefont {R.}~\bibnamefont {Podgornik}},\
  }\href@noop {} {\bibfield  {journal} {\bibinfo  {journal} {EPL (Europhysics
  Letters)}\ }\textbf {\bibinfo {volume} {113}},\ \bibinfo {pages} {26004}
  (\bibinfo {year} {2016})}\BibitemShut {NoStop}%
\bibitem [{\citenamefont {Moreira}\ and\ \citenamefont
  {Netz}(2000)}]{moreira2000strong}%
  \BibitemOpen
  \bibfield  {author} {\bibinfo {author} {\bibfnamefont {A.~G.}\ \bibnamefont
  {Moreira}}\ and\ \bibinfo {author} {\bibfnamefont {R.~R.}\ \bibnamefont
  {Netz}},\ }\href@noop {} {\bibfield  {journal} {\bibinfo  {journal} {EPL
  (Europhysics Letters)}\ }\textbf {\bibinfo {volume} {52}},\ \bibinfo {pages}
  {705} (\bibinfo {year} {2000})}\BibitemShut {NoStop}%
\bibitem [{\citenamefont {Netz}(2001)}]{netz2001electrostatistics}%
  \BibitemOpen
  \bibfield  {author} {\bibinfo {author} {\bibfnamefont {R.~R.}\ \bibnamefont
  {Netz}},\ }\href@noop {} {\bibfield  {journal} {\bibinfo  {journal} {The
  European Physical Journal E}\ }\textbf {\bibinfo {volume} {5}},\ \bibinfo
  {pages} {557} (\bibinfo {year} {2001})}\BibitemShut {NoStop}%
\bibitem [{\citenamefont {{\v{S}}amaj}\ and\ \citenamefont
  {Trizac}(2011)}]{vsamaj2011wigner}%
  \BibitemOpen
  \bibfield  {author} {\bibinfo {author} {\bibfnamefont {L.}~\bibnamefont
  {{\v{S}}amaj}}\ and\ \bibinfo {author} {\bibfnamefont {E.}~\bibnamefont
  {Trizac}},\ }\href@noop {} {\bibfield  {journal} {\bibinfo  {journal}
  {Physical Review E}\ }\textbf {\bibinfo {volume} {84}},\ \bibinfo {pages}
  {041401} (\bibinfo {year} {2011})}\BibitemShut {NoStop}%
\bibitem [{\citenamefont {Hastings}\ and\ \citenamefont
  {Levitov}(1996)}]{hastings1996laplacian}%
  \BibitemOpen
  \bibfield  {author} {\bibinfo {author} {\bibfnamefont {M.~B.}\ \bibnamefont
  {Hastings}}\ and\ \bibinfo {author} {\bibfnamefont {L.~S.}\ \bibnamefont
  {Levitov}},\ }\href@noop {} {\bibfield  {journal} {\bibinfo  {journal} {arXiv
  preprint cond-mat/9607021}\ } (\bibinfo {year} {1996})}\BibitemShut {NoStop}%
\bibitem [{\citenamefont {Mullins}\ and\ \citenamefont
  {Sekerka}(1963)}]{mullins1963morphological}%
  \BibitemOpen
  \bibfield  {author} {\bibinfo {author} {\bibfnamefont {W.~W.}\ \bibnamefont
  {Mullins}}\ and\ \bibinfo {author} {\bibfnamefont {R.~F.}\ \bibnamefont
  {Sekerka}},\ }\href@noop {} {\bibfield  {journal} {\bibinfo  {journal}
  {Journal of applied physics}\ }\textbf {\bibinfo {volume} {34}},\ \bibinfo
  {pages} {323} (\bibinfo {year} {1963})}\BibitemShut {NoStop}%
\bibitem [{\citenamefont {Mullins}\ and\ \citenamefont
  {Sekerka}(1988)}]{mullins1988stability}%
  \BibitemOpen
  \bibfield  {author} {\bibinfo {author} {\bibfnamefont {W.~W.}\ \bibnamefont
  {Mullins}}\ and\ \bibinfo {author} {\bibfnamefont {R.}~\bibnamefont
  {Sekerka}},\ }in\ \href@noop {} {\emph {\bibinfo {booktitle} {Dynamics of
  Curved Fronts}}}\ (\bibinfo  {publisher} {Elsevier},\ \bibinfo {year}
  {1988})\ pp.\ \bibinfo {pages} {345--352}\BibitemShut {NoStop}%
\bibitem [{\citenamefont {Bazant}\ \emph {et~al.}(2003)\citenamefont {Bazant},
  \citenamefont {Choi},\ and\ \citenamefont
  {Davidovitch}}]{bazant2003dynamics}%
  \BibitemOpen
  \bibfield  {author} {\bibinfo {author} {\bibfnamefont {M.~Z.}\ \bibnamefont
  {Bazant}}, \bibinfo {author} {\bibfnamefont {J.}~\bibnamefont {Choi}}, \ and\
  \bibinfo {author} {\bibfnamefont {B.}~\bibnamefont {Davidovitch}},\
  }\href@noop {} {\bibfield  {journal} {\bibinfo  {journal} {Physical review
  letters}\ }\textbf {\bibinfo {volume} {91}},\ \bibinfo {pages} {045503}
  (\bibinfo {year} {2003})}\BibitemShut {NoStop}%
\bibitem [{\citenamefont {Mirzadeh}\ and\ \citenamefont
  {Bazant}(2017)}]{mirzadeh2017electrokinetic}%
  \BibitemOpen
  \bibfield  {author} {\bibinfo {author} {\bibfnamefont {M.}~\bibnamefont
  {Mirzadeh}}\ and\ \bibinfo {author} {\bibfnamefont {M.~Z.}\ \bibnamefont
  {Bazant}},\ }\href@noop {} {\bibfield  {journal} {\bibinfo  {journal}
  {Physical review letters}\ }\textbf {\bibinfo {volume} {119}},\ \bibinfo
  {pages} {174501} (\bibinfo {year} {2017})}\BibitemShut {NoStop}%
\bibitem [{\citenamefont {Dowell}\ and\ \citenamefont
  {Rinfret}(1960)}]{dowell1960low}%
  \BibitemOpen
  \bibfield  {author} {\bibinfo {author} {\bibfnamefont {L.~G.}\ \bibnamefont
  {Dowell}}\ and\ \bibinfo {author} {\bibfnamefont {A.~P.}\ \bibnamefont
  {Rinfret}},\ }\href@noop {} {\bibfield  {journal} {\bibinfo  {journal}
  {Nature}\ }\textbf {\bibinfo {volume} {188}},\ \bibinfo {pages} {1144}
  (\bibinfo {year} {1960})}\BibitemShut {NoStop}%
\bibitem [{\citenamefont {Steytler}\ \emph {et~al.}(1983)\citenamefont
  {Steytler}, \citenamefont {Dore},\ and\ \citenamefont
  {Wright}}]{steytler1983neutron}%
  \BibitemOpen
  \bibfield  {author} {\bibinfo {author} {\bibfnamefont {D.}~\bibnamefont
  {Steytler}}, \bibinfo {author} {\bibfnamefont {J.}~\bibnamefont {Dore}}, \
  and\ \bibinfo {author} {\bibfnamefont {C.}~\bibnamefont {Wright}},\
  }\href@noop {} {\bibfield  {journal} {\bibinfo  {journal} {The Journal of
  Physical Chemistry}\ }\textbf {\bibinfo {volume} {87}},\ \bibinfo {pages}
  {2458} (\bibinfo {year} {1983})}\BibitemShut {NoStop}%
\bibitem [{\citenamefont {Mayer}\ and\ \citenamefont
  {Hallbrucker}(1987)}]{mayer1987cubic}%
  \BibitemOpen
  \bibfield  {author} {\bibinfo {author} {\bibfnamefont {E.}~\bibnamefont
  {Mayer}}\ and\ \bibinfo {author} {\bibfnamefont {A.}~\bibnamefont
  {Hallbrucker}},\ }\href@noop {} {\bibfield  {journal} {\bibinfo  {journal}
  {Nature}\ }\textbf {\bibinfo {volume} {325}},\ \bibinfo {pages} {601}
  (\bibinfo {year} {1987})}\BibitemShut {NoStop}%
\bibitem [{\citenamefont {Murray}\ \emph {et~al.}(2005)\citenamefont {Murray},
  \citenamefont {Knopf},\ and\ \citenamefont {Bertram}}]{murray2005formation}%
  \BibitemOpen
  \bibfield  {author} {\bibinfo {author} {\bibfnamefont {B.~J.}\ \bibnamefont
  {Murray}}, \bibinfo {author} {\bibfnamefont {D.~A.}\ \bibnamefont {Knopf}}, \
  and\ \bibinfo {author} {\bibfnamefont {A.~K.}\ \bibnamefont {Bertram}},\
  }\href@noop {} {\bibfield  {journal} {\bibinfo  {journal} {Nature}\ }\textbf
  {\bibinfo {volume} {434}},\ \bibinfo {pages} {202} (\bibinfo {year}
  {2005})}\BibitemShut {NoStop}%
\bibitem [{\citenamefont {Moore}\ and\ \citenamefont
  {Molinero}(2011)}]{moore2011cubic}%
  \BibitemOpen
  \bibfield  {author} {\bibinfo {author} {\bibfnamefont {E.~B.}\ \bibnamefont
  {Moore}}\ and\ \bibinfo {author} {\bibfnamefont {V.}~\bibnamefont
  {Molinero}},\ }\href@noop {} {\bibfield  {journal} {\bibinfo  {journal}
  {Physical Chemistry Chemical Physics}\ }\textbf {\bibinfo {volume} {13}},\
  \bibinfo {pages} {20008} (\bibinfo {year} {2011})}\BibitemShut {NoStop}%
\bibitem [{\citenamefont {Finney}\ \emph {et~al.}(2002)\citenamefont {Finney},
  \citenamefont {Hallbrucker}, \citenamefont {Kohl}, \citenamefont {Soper},\
  and\ \citenamefont {Bowron}}]{finney2002structures}%
  \BibitemOpen
  \bibfield  {author} {\bibinfo {author} {\bibfnamefont {J.}~\bibnamefont
  {Finney}}, \bibinfo {author} {\bibfnamefont {A.}~\bibnamefont {Hallbrucker}},
  \bibinfo {author} {\bibfnamefont {I.}~\bibnamefont {Kohl}}, \bibinfo {author}
  {\bibfnamefont {A.}~\bibnamefont {Soper}}, \ and\ \bibinfo {author}
  {\bibfnamefont {D.}~\bibnamefont {Bowron}},\ }\href@noop {} {\bibfield
  {journal} {\bibinfo  {journal} {Physical review letters}\ }\textbf {\bibinfo
  {volume} {88}},\ \bibinfo {pages} {225503} (\bibinfo {year}
  {2002})}\BibitemShut {NoStop}%
\end{thebibliography}%

\end{document}